\documentclass[aps,pra, onecolumn,14 pts,a4paper,showpacs]{revtex4}

\usepackage{amssymb,amsmath,amsfonts,graphicx}
\usepackage{bm}

\begin{document}

\title{Prediction of catastrophes: an experimental model}

\author{Randall D. Peters$^1$, Martine Le Berre$^2$  and Yves Pomeau $^3$}

\affiliation{$^1$ Department of Physics, Mercer University, Macon, Georgia, USA.
\\ $^2$ Institut des Sciences Mol\'eculaires d'Orsay ISMO-CNRS, Univ. Paris-Sud, Bat. 210, 91405 Orsay, France.
\\$^3$ Department of Mathematics, University of Arizona, Tucson, USA.}

\date{\today }

\begin{abstract}
\begin{center}
\end{center}

Catastrophes of all kinds
can be roughly defined as short duration-large amplitude events following and followed by long periods of ``ripening''.
Major earthquakes surely belong to the class of `catastrophic' events. Because of the space-time scales involved, an
experimental approach is often difficult, not to say impossible, however desirable it could be.
Described in this article is a ``laboratory'' setup that yields data of a type that is amenable to theoretical
methods of prediction.  Observations are made of a critical slowing down in the noisy signal of a solder wire creeping
under constant stress. This effect is shown to be a fair signal of the forthcoming catastrophe in both of two dynamical models.
The first is an ``abstract'' model
in which a time dependent quantity drifts slowly but makes quick jumps from time to time.  The second is a
realistic physical model for the collective motion of dislocations (the Ananthakrishna set of equations for creep).
Hope thus exists that similar changes in the response to noise could forewarn catastrophes in other situations,
where such
 precursor effects should manifest
 early enough.

\end{abstract}

\maketitle

\section{Introduction}

Catastrophes as defined in the abstract are related to a class of phenomena sometimes called ``relaxation'' oscillations. The
observation of two widely separated time scales
makes relaxation oscillations a good a priori subject of theoretical investigation, because one may suspect that their
formation results from the existence of a (more or less hidden) small parameter.  This gives some hope of a ``general'' theory
based upon the small size of the parameter. To take an example, some major earthquakes lasting a few tens of seconds occur
in the same general area about every hundred years.  They thus involve a ratio of `typical' times in the neighborhood of $10^{-9}$,
which is a very small number.
In the present study a laboratory experiment was devised which displays similar relaxation oscillations,
but acting time-wise on a scale that is convenient for investigation.  Experimental observations are explained
in light of a two-part \textit{dynamical bifurcation} model of catastrophe.
 Comprising
the local form of a physical model shown to be valid
 when close to the catastrophe, the striking result is as follows.
 In response to an external source of
noise the two models (the local one and the physical one) predict fluctuations with a correlation time that increases
before the catastrophe, just as is observed experimentally.
The important point is that this well known critical slowing down phenomenon occurs significantly {\it{before}}
the transition and could be used to forewarn it.

 We study the creeping of a soft metal under constant stress. Accurate time records show that this creeping actually displays
the following time dependent noisy component. The wire typically lengthens slowly with background (non-thermal) noise, with
sometimes a ``large'' sliding event, followed again by a noisy slow lengthening regime, etc.

Plastic deformation of solids is a complex phenomenon, not yet fully understood \cite{AnantaR}. It has long been observed
to take place in a non-smooth manner. Most studies have focused on the Portevin-Le Chatelier effect observed under constant strain
rate conditions, whereas the present experiment is concerned with creep at constant stress. In both
cases large steps (cyclic slips) are embedded in a noisy
background. There is a general agreement that this is due to the complex dynamics of networks of dislocations whose motion
is a means for the stressed solid to flow. As reported in section \ref{sec:experiment} below, careful observations of the creep
in strained Sn-Pb quasi-eutectic material (a solder wire) show the following :

{\it{i)}} On average a sample under constant stress lengthens at constant rate.

{\it{ii)}} Continuous monitoring shows time dependent fluctuations of this length superposed on its secular increase.

{\it{iii)}} From time to time the length jumps by steps. Afterwards a noisy and steady (on average) length increase is
recovered until the next jump, etc.

Explaining all this remains a challenge for the common models of creep. Nevertheless it is of great interest because it
can be seen as a laboratory model of other far less accessible phenomena like earthquakes, where on average there is also
a continuous slow sliding with random microseismic noise, interrupted by large fast sliding steps characterizing
major earthquakes.
We recently introduced the idea  \cite{nous-saddle} that in such systems the slow to fast transition can be described by
a saddle-node bifurcation in a dynamical system evolving slowly with time. We pointed out the interest in such a modelization,
which allows one to theoretically predict the response of this dynamical system to a noise source.  It was shown that the
induced fluctuations drift toward low frequencies (i.e., toward large correlation times) {\it{before}} the
transition, and so could be used as a forewarning.

Here we show that the dynamical saddle-node bifurcation model is the reduced form of a set of equations previously used
to describe the Portevin-Le Chatelier effect in metals, or metallic alloys, i.e., the Ananthakrishna (AK) model \cite{AnantaR}.
The saddle-node model describes fast transitions
resulting from the intrinsic dynamics of the original system which can be described locally (close to the step) as a slowly
rocking potential system. Using the AK model as well as the saddle-node reduced model we show that, with an added external
source of noise, the correlation time of the fluctuations increases before the transition, following the classical scenario
of slowing down at bifurcation points.  As was shown long ago by
Dorodnytsin \cite{dor}, the same local dynamics describe the transition from slow manifold to fast transients in
relaxation oscillations of dynamical models, like the van der Pol equation in the strongly nonlinear limit.
By looking at the fluctuations of the length of our samples, we found this behavior near the step-like transitions,
with the characteristic drift to low frequencies before the transition.

Unlike recent publications that have presented the idea \cite{russe}
that precursors of
earthquakes could be found in the mechanical response to external perturbations (such as the increase of the fluctuations and
their slowing down near transitions) our idea goes further, by giving an order of magnitude of the precursor time.
The authors of \cite{russe} do not introduce the effect of a given time dependence of the parameters and consider only
systems with steady parameter values on both sides of the bifurcation.
To quote reference \cite{russe1} ``The suggested approach to analytical study of any kind of catastrophes
is based essentially on the solution of a stationary problem
of the possibility and conditions of the unstable equilibrium
state in the system in question''.
Without taking into account explicitly the time dependence of the parameters sweeping the bifurcation set,
it is impossible to get the time scale for predictions.  As we show, this scale depends crucially on the rate of
change of the parameters near the bifurcation, which may be estimated from the knowledge of the ratio of the two time scales (fast and slow ones) for saddle-node models.

 An important challenge lies in the difficulty of stating a suitable model for a given catastrophic event. Actually there is
more than one class of possible slow-to-fast transitions in dynamical systems. The dynamical saddle-node can be seen as
belonging to the class of systems with an equilibrium point losing stability as a parameter changes.  It is not
a loss of stability, but rather a loss of existence of the equilibrium point, occurring at the folding point of the slow manifold. However slow-to-fast transition
may happen without any folding of the slow manifold. As shown in \cite{Chua} exploring  a very often used mathematical representation of stick-slip behavior, the Dieterich-Ruina equations, the slow-fast transition
 can originate from the finite time singularity of the
slow dynamics itself. In that case, critical speed-up, or drift of fluctuations toward large frequencies, is found to replace the critical slowing-down effect.
This nonlinear phenomenon is obviously outside the class of phenomena explainable by a stability
analysis of equilibria of dynamical systems. We refer the interested reader to the paper on this subject \cite{Chua}.
In the present paper we do not consider this case, because it is clearly not the one
observed in the creeping experiments.
 Note that at this time it is unknown if real earthquakes
(as well as other observed catastrophes) belong to the saddle-node case with a slowing-down near the transition or to the finite
time singularity case with speed-up expected.

The creeping experiment is described in section \ref{sec:experiment} together with the striking spectral observations. In
the two sections following section \ref{sec:experiment}, we present our theoretical approach. We  introduce in
section \ref{sec:dynsaddbif} an abstract dynamical model showing fast jumps. In this model, one assumes that, as a
dynamical system, the jump follows a ``saddle-node'' bifurcation. There a pair of fixed points, one  locally stable, the
other locally unstable, merge and disappear as a control parameter changes. Afterwards, the system moves quickly to a
new equilibrium state that is
at finite distance (in phase space)
from previous equilibria, whence the jump.
After reviewing the saddle-node bifurcation in this light, we assume that the
 parameter changes with time, namely that the parameter defining the bifurcation is itself a slow function of time. When, by this change, the parameter
crosses the bifurcation value, the dynamical system makes an abrupt transition and jumps ``generically'' from one equilibrium state
to another.
In section \ref{sec:AK} we show that this ``abstract'' model is pertinent for describing the slow-fast transition in the
relaxation regime of a set of equations derived by Ananthakrishna for creeping in the relaxation regime.
This set of equations describes the dynamics of populations of dislocations in the creeping solid. The model shows
relaxation oscillations in which slow drift is interrupted by fast variations. Near the slow-fast transition, we show
that this model reduces to the generic equation quoted above, for a certain range of parameters. Therefore it displays
a typical critical slowing-down in its response to external noise, in agreement with the real data reported in
section \ref{sec:experiment}. Of particular interest is the fact that, from the experimental data, one can predict
in advance a ``large slip'' event.  The event is preceded by a shift toward low frequencies in the random fluctuations
of specimen length.

\section{ Creeping experiment}
\label{sec:experiment}


Choose a uniform``wire'' that creeps under small stress at room temperature, fabricated from a much studied
material, Sn-Pb solder, alloyed to be nearly eutectic. To establish creep at virtually constant stress,
let one end of the wire be fixed and at the other end establish a constant force of tension.
By this means we found the length of the wire to increase on average at constant rate.
By highly accurate monitoring of this length vs time
we observed a small time dependent part with the following pattern: on
a background of fluctuations, from time to time a
large slip event was observed,
after which the continuous lengthening with a small background noise was
recovered.
The observed noise is non-thermal, since thermal noise has a far too small amplitude to be relevant.
It is assumed to result from rearrangement of defect structures in the
poly-crystalline structure of the quasi-eutectic, involving dislocation dynamics. We
assume that this noise originates from a source that is to first-order independent of the overall
lengthening of the wire. This is equivalent to saying that it is due to
ongoing micro-scale events triggered by the imposed stress, independent
of the global creeping. Therefore we analyzed the
response to this noise source according to the AK
equations subject to an external noise (see below).

\subsection{apparatus}
\begin{figure}[htbp]
\centerline{$\;\;$
\includegraphics[height=2in]{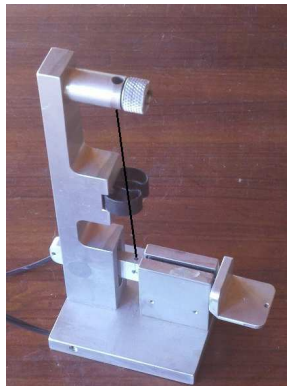}
$\;\;$}
\caption{Extensometer used in the experiment.
}
\label{Fig:solder}
\end{figure}

The instrument used in these experiments, which is pictured in Fig.\ref{Fig:solder}, is an extensometer
\cite{apparatus1}.
 Young's modulus can be accurately measured with a wire specimen, by
placing different size masses on the weight pan. The trace of the vertical wire holding up the boom/weight pan
in Fig.\ref{Fig:solder} has been enforced (colored in black) to be visible in the image. Though not presently used, the
black clamp was for purpose of holding a power resistor that was employed to measure the specimen's thermal coefficient of expansion.
To measure temperature
changes, a solid state thermometer was placed down into the sample space, through the hole seen
near the top knurled clamp.

To calibrate the instrument a tungsten wire of diameter 0.1 mm was mounted in the extensometer, and signal output level
changes were recorded as various gram-mass
standards were placed on the weight pan.  By using the known Young's
modulus for tungsten, the resulting measurements yielded a constant of 1.0 nm per analog to digital count,
for the 24 bit adc employed, which is sold by Symmetric Research \cite{apparatus4}.  This constant is
applicable to the data presently reported. A different measurement technique yielded essentially the same calibration constant.
A He-Ne laser was used with a mirror, operating as an optical lever, to measure boom position change as different masses
were placed on the pan.

On the basis of two factors, ordinary solder was chosen for the present study.  First of all, an unusual property of tin is well
known, when an ingot of the metal is strained by large amounts.  The sound which it then emits is well described
by the German word ``zinngeschrei", which translates ``tin cries''.

The initiative for this work was also influenced by the observation of unusual spectral features in the output from a
novel seismograph \cite{apparatus5}.
The unusual low-frequency motions of the Earth's crust that were then observed to precede an earthquake \cite{apparatus6}
are readily seen by the VolksMeter.
This is due to the instrument's use of a ``displacement'' sensor, rather than the ``velocity'' sensor used by
conventional seismometers.
It was therefore natural to consider an alloy of tin, with the expectation that its defect properties should be
more like those of the earth than is possible for a pure metal. The tin alloy for our study was ordinary soft solder (60\% Sn/
40\% Pb), used universally in the electronics industry. Although the pure `Eutectic' alloy is actually 63\% Sn/37\% Pb, we will
nevertheless use this word to describe our specimen in the discussions that follow.

    The stress level due to the load placed on the Pb-Sn wire used in the present experiments was considerably smaller than the
typically $10 MPa$ used in usual creep studies. The present load was due solely to the weight of aluminum comprising the
boom plus empty weight pan of the extensometer.  This stress value was estimated to be $0.5 MPa$, based on considerations that
include the distance of the wire's attachment point from the position of the (fine) roller bearing in the
unpright housing, that supports the boom on its end opposite the pan.

	  \begin{figure}[htbp]
\centerline{$\;\;$
 (a)\includegraphics[height=1.5in]{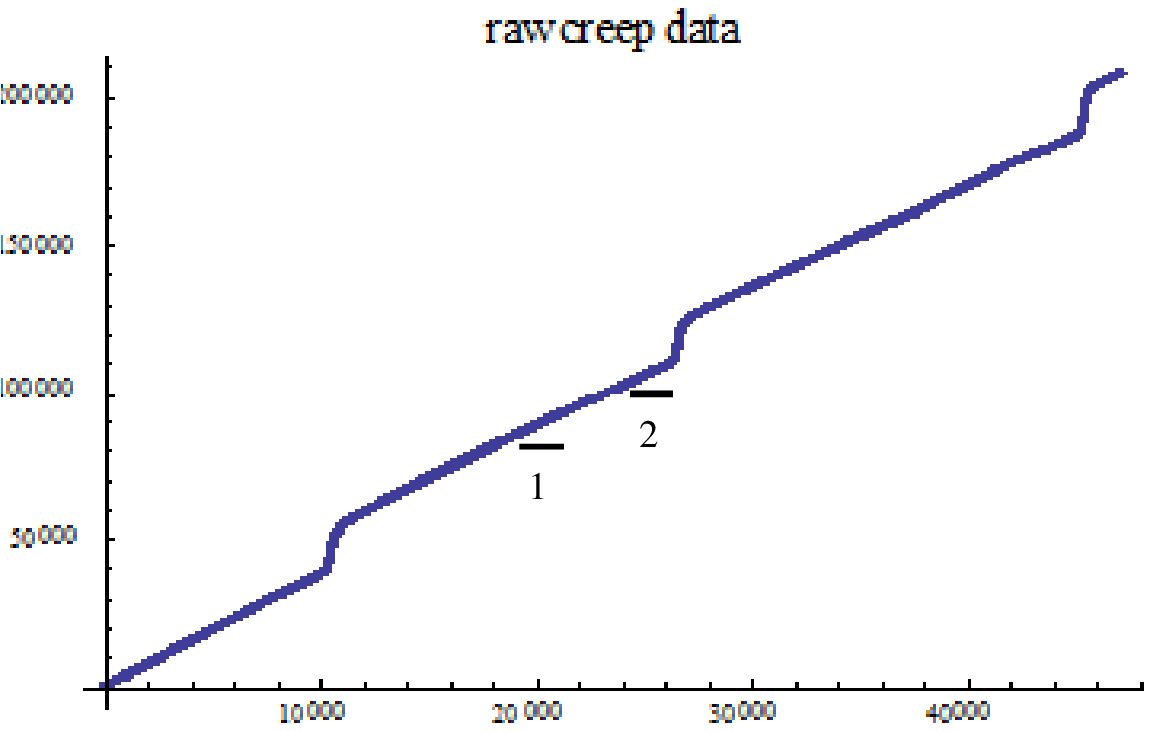}
 (b)\includegraphics[height=1.5in]{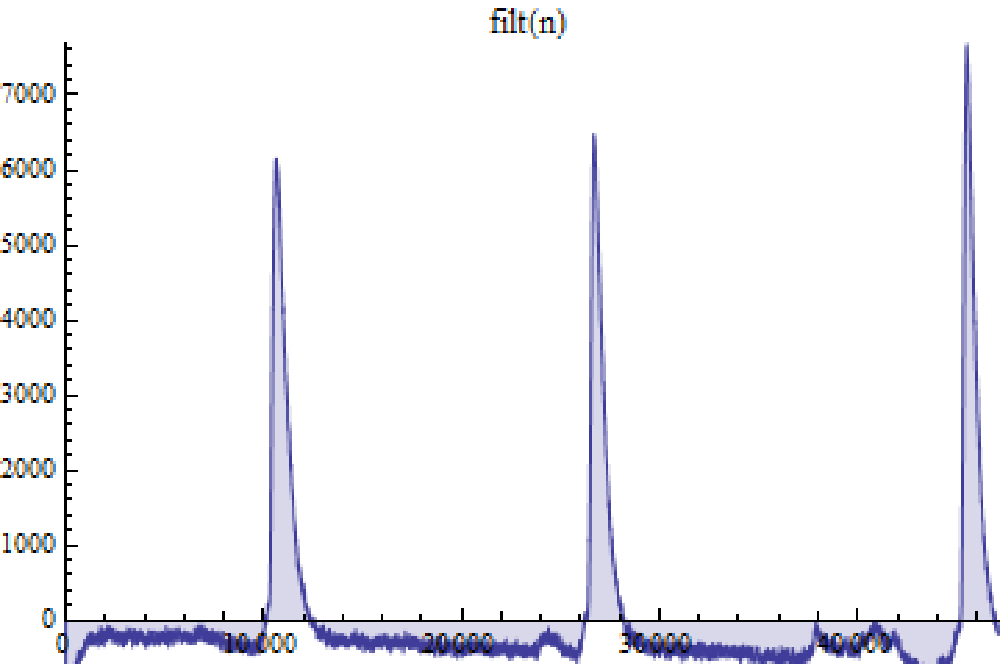}
$\;\;$}
\caption{(a) Elongation of the wire versus time, raw signal $x(n)$. (b) Filtered signal  $y(n)$  at the exit of the single pole high-pass filter (\ref{eq:filt}).
}
\label{Fig:creep-exp}
\end{figure}

During a typical avalanche, the elongation of a $13.5 cm$ long, $1.6 mm$ diameter specimen, would be about $30 \mu m$.  Before the
avalanche, typical lengthening velocity is $5\mu m/s$ and typical rms fluctuations observed in the wire length
(with secular term removed by high-pass filtering) would be about $50 nm$.

An example of creep record (length growth versus time) is shown in Fig.\ref{Fig:creep-exp}(a). The fluctuations of the raw
signal are not visible because they are much smaller than the average length variations.
\subsection{Data analysis}
To eliminate the average growth, we use a standard technique of filtering. At the exit of a low-pass single pole filter,
the filtered signal $y(n)$ is given by the recursive formula

\begin{equation}
y_n = \frac{1+p}{2}(x_n-x_{n-1})+py_{n-1}
\mathrm{,}
\label{eq:filt}
\end{equation}

where $x(n)$ stands for the raw creep signal. In other words the filtered signal is the convolution product of the
raw signal derivative by an exponential response function $R(t)=\exp{-2\pi f_c t}$, where $f_c$ is the corner
frequency of the filter. In equation (\ref{eq:filt}) the parameter $p$ is given by $p=\exp{-2\pi f_c \delta t}$
where  $\delta t $ is the sampling time of the record, equal to $\frac{1}{130} sec$ in the experiment. Using a
corner frequency $f_c=50 mHz$, the filtered signal takes the form shown in Fig.\ref{Fig:creep-exp}(b).
	  \begin{figure}[htbp]
\centerline{$\;\;$
 (a)\includegraphics[height=1.2in]{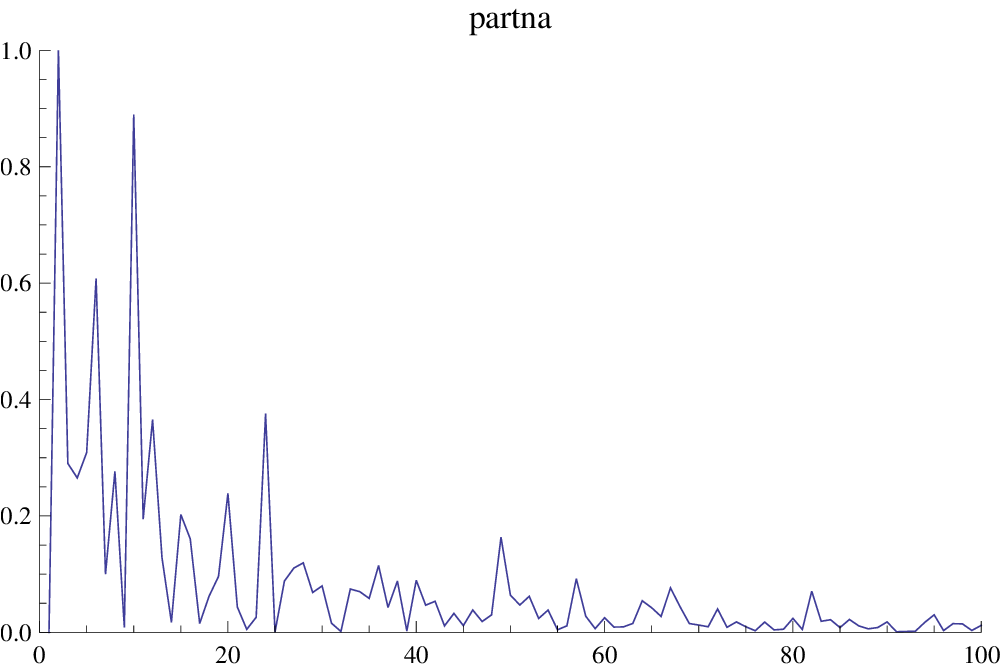}
 (b)\includegraphics[height=1.2in]{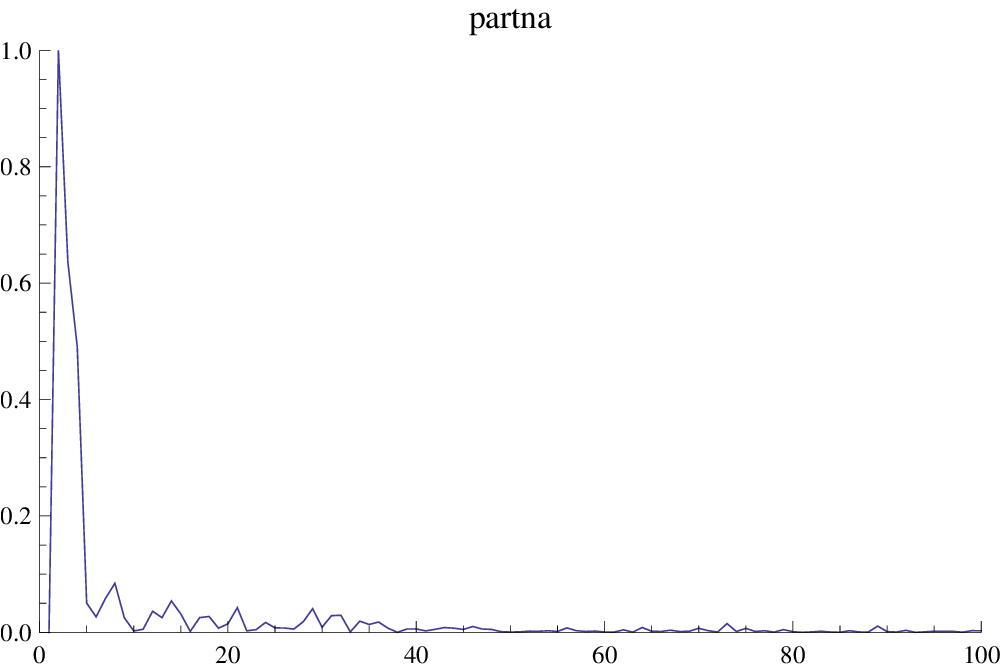}
  $\;\;$}
  \centerline{$\;\;$
 (c)\includegraphics[height=1.2in]{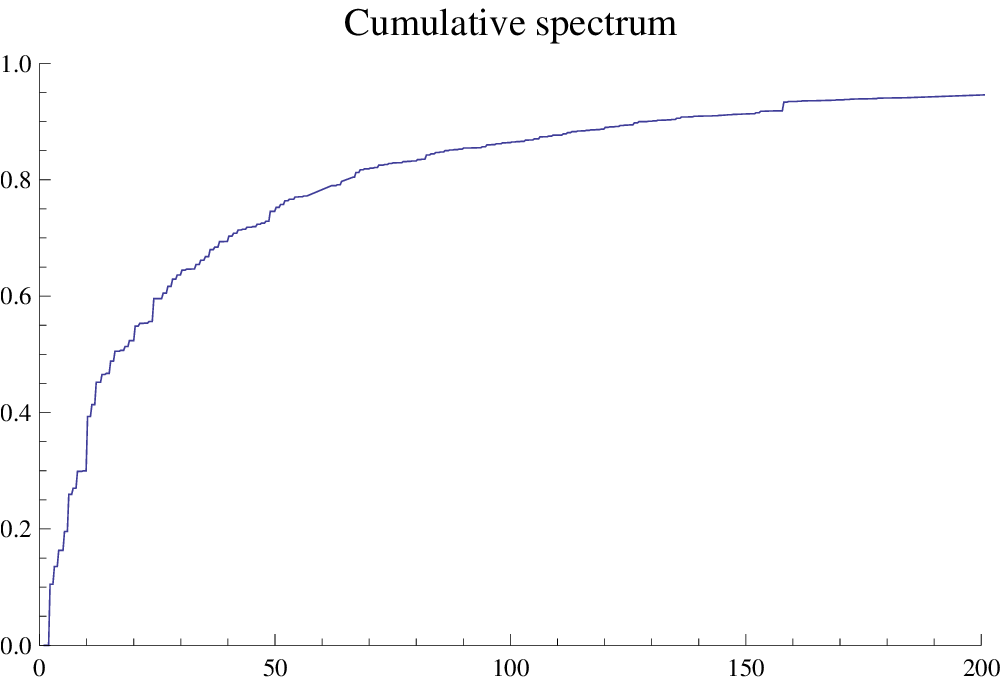}
 (d)\includegraphics[height=1.2in]{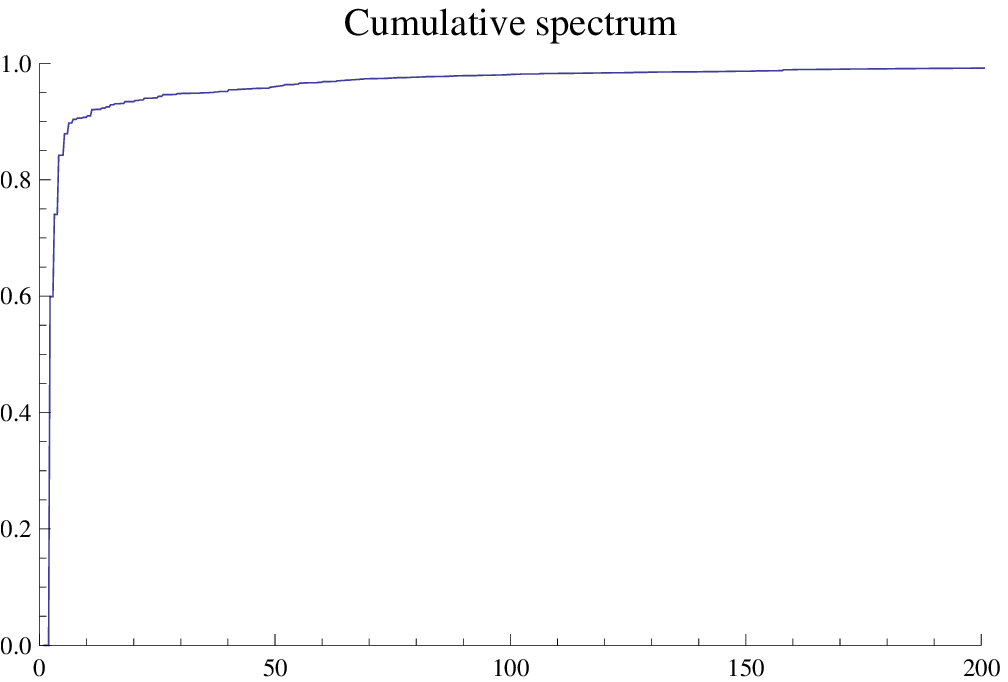}
  $\;\;$}
   \centerline{$\;\;$
 (e)\includegraphics[height=1.2in]{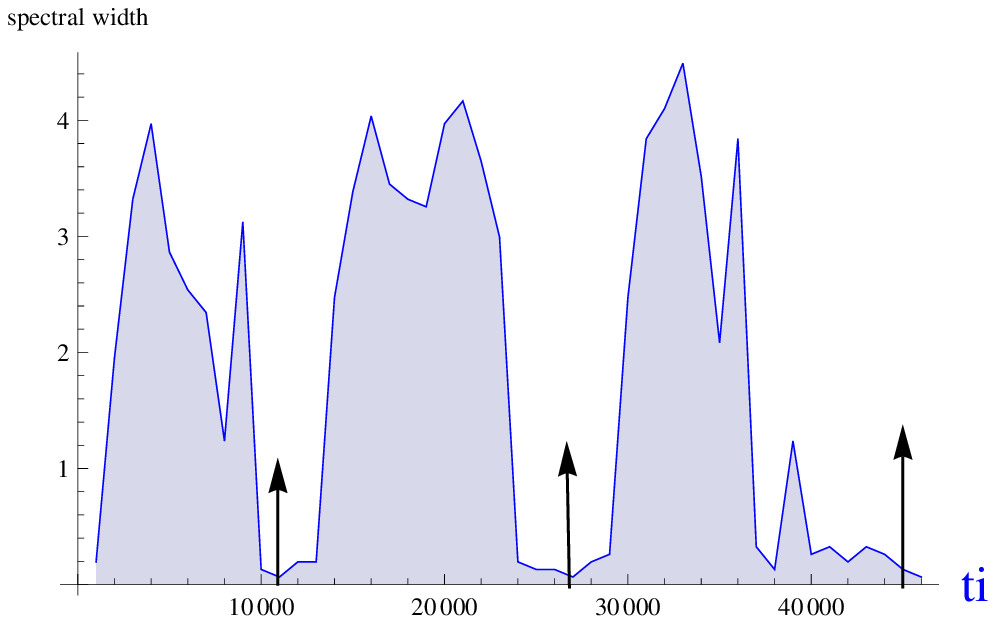}
  $\;\;$}
\caption{
(a) and (b) Normalized spectral density of $y(n)$ before the second step (the step starts at $n=26370$), for the two
time intervals (1-2) indicated in Figure (\ref{Fig:creep-exp})-a, at the abscissa  $18370<n<20370$ and $24370<n=26370$
respectively, each for a time interval of $N=2000$ counts.
(c-d) Cumulative spectra corresponding to figures (a-b) respectively
(e)Width $w(t)$ of the  cumulative spectrum versus time (at height equal to $75$ per cent of
the maximum), the vertical arrows indicate the very beginning of the fast steps.
}

\label{Fig:creep-exp2}
\end{figure}
In order to analyze the spectral properties of the fluctuations, and see how they evolve with time, we must consider separate sequences of a given time duration.
One significant difficulty in this
analysis is the proper choice for the magnitude of this time duration. The window cannot be too large, lest the spectrum becomes invariant with time.  Conversely, if the window is too narrow, there will be excessive
noise in the spectral (or correlation) signal. We believe the
time windows have been judiciously chosen; so that there is a significant increase of the correlation
time before a ``burst''.
Such increase of the correlation time (actually the decrease of
the frequency width of the signal) was found to happen before every
sliding event, giving some hope that this ``critical slowing down" is
pertinent for predicting the ``catastrophe" before it occurs.
For the present work a time duration of order $1/10~th$ of the interval between adjacent bursts was chosen. The spectral density of a given sequence ($n_i,n_i+N$) of the filtered signal

\begin{equation}
S(n) = |\frac{1}{\sqrt{N}}\sum_{n_i}^{n_i+N} y(k)\exp{2i\pi nk}|^2
\mathrm{,}
\label{eq:spec}
\end{equation}

 changes significantly during the creep process. The striking effect is the shift of the  spectral density toward low frequencies in the last stage of the slow regime, i.e., just before the burst. Examples of this phenomenon are presented in the figures (\ref{Fig:creep-exp2})(a-b), which show the spectra corresponding to the time intervals ($1,2)$ indicated in Figure (\ref{Fig:creep-exp})-a. The two spectra clearly differ. The first spectral density (a) displays a large number of components; whereas the second spectral density, corresponding to the time interval just before the burst (figure b), is  concentrated close to zero frequency.
To quantify this effect, we calculate the cumulative spectrum

\begin{equation}
C_S(n) = \sum_1^n S(j)
\mathrm{,}
\label{eq:cumul}
\end{equation}
which is a smooth curve whose asymptotic value (for n=N) gives the experimental variance of the fluctuations during the sequence considered

\begin{equation}
\sigma^2 = C_S(N)
\mathrm{.}
\label{eq:sigma2}
\end{equation}

 \begin{figure}[htbp]
\centerline{$\;\;$
(a)\includegraphics[height=1.2in]{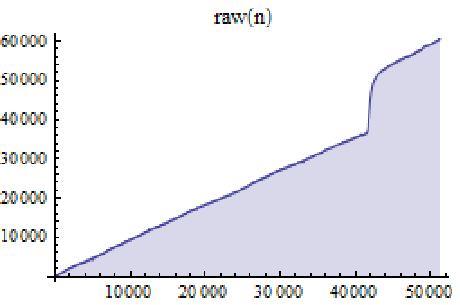}
(b)\includegraphics[height=1.2in]{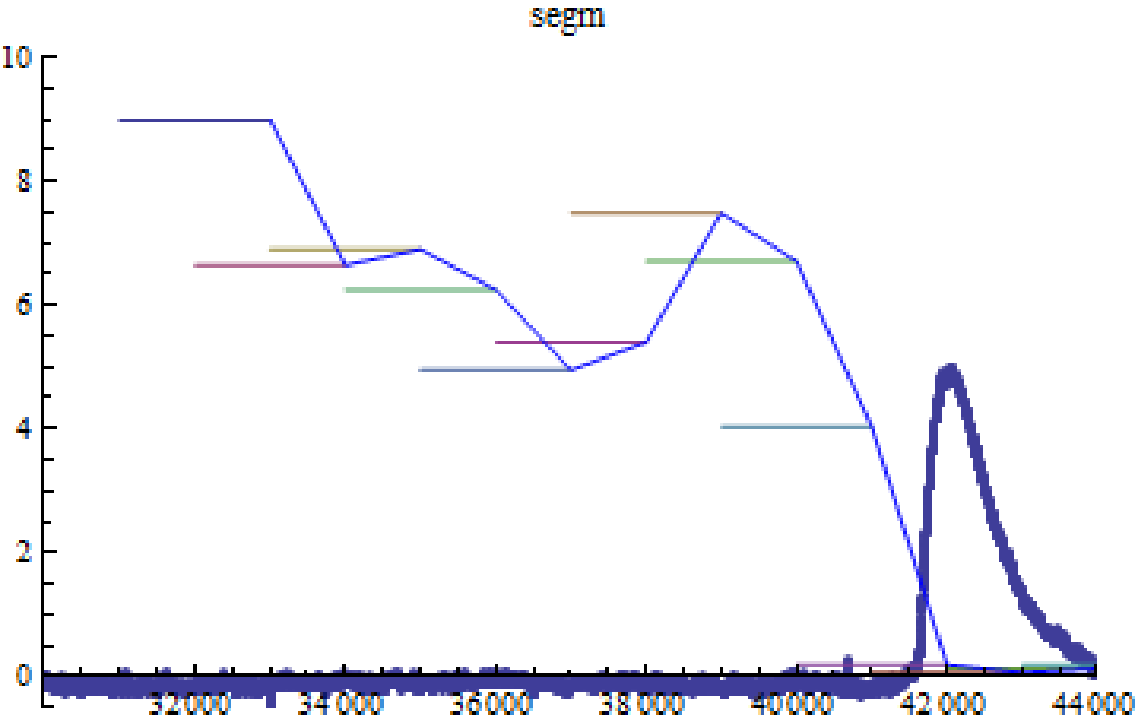}
 $\;\;$}
\caption{ (a) Creep data versus time, (b) Width of the cumulative spectra before the burst with $N=2000$. The horizontal segments ending at each point indicates the time interval (before this point) over which the width is calculated.}
\label{Fig:creep-exp3}
\end{figure}

The shift of the density spectrum toward low frequencies is visually more clear in its associated cumulative spectrum, as seen in figures (c-d).  We define the ``low frequency extension"  $w$ of a sequence by the width of
the cumulative spectrum corresponding to  $75$ per cent of its maximum value,

\begin{equation}
C_S(w)= \frac{3}{4} C_S(N)
\mathrm{.}
\label{eq:cumwidth}
\end{equation}

Shown in figure 3(e) is the evolution of the spectral width for the whole experimental record. A drop in $w$ is clearly seen to occur before each burst.

The important point is that the decrease of the width before the burst is experimentally foreseeable because it occurs during a ``precursor" time which is noticeably larger than the duration of a sequence $N \delta t$. This is illustrated in Fig.\ref{Fig:creep-exp3}(b) which displays a zoom of the spectral width evolution before the burst.

The two following sections are devoted to theoretical models, the first one shows the main features of what is measured in the experiment on creeping, namely the critical slowing down of the fluctuation spectrum occurring before the sliding event, the second one is a model of creeping found in the literature which consists in a set of nonlinear coupled ordinary differential equations, derived by Ananthakrishna  and collaborators \cite{AK2}.

\section{Dynamical saddle-node bifurcation}
\label{sec:dynsaddbif}

This section explains why a saddle-node bifurcation with a slow sweeping of the bifurcation parameter exhibits a slowing down in its response to a source of noise in a window of time extending well before the bifurcation itself. This ``abstract' model has no direct connection with the physical phenomenon of creeping. In the section afterwards, however, we explain that a model of creeping shows the same slowing down in a range of parameters, linked to a local saddle-node bifurcation.

Consider first the saddle-node bifurcation of a ``gradient flow", that is a damped dynamical system such that a coordinate $x(t)$ is a solution of the equation of motion of the form

\begin{equation}
\frac{{\mathrm{d}}x}{{\mathrm{d}}t} = -\frac{\partial V}{\partial x}
\mathrm{.}
\label{eq:grad}
\end{equation}

In this equation $V(x)$ is a potential, and the dynamics tends to everywhere lower the value of $V(x)$.

 The catastrophe theory of Thom and Arnol'd \cite{cat} studies how steady equilibria of  equations like (\ref{eq:grad}) change under smooth deformations of the potential $V(x)$. Below we consider a different kind of question, namely what happens to the solution of equation (\ref{eq:grad})  when the potential $V(x)$ becomes itself a slowly varying function of time, and particularly when a pair of equilibrium points disappears by a saddle-node bifurcation. Indeed this question of the sweeping across bifurcations has been already widely studied \cite{sweep} with various applications in mind (by "sweeping" we mean crossing of a transition point with a time dependent parameter in the equation(s) of motion). However, to the best of our knowledge the occurrence of an intermediate time scale in the case of slow sweeping has been overlooked, although we believe it to be crucial for a strategy of foretelling catastrophes in the real world.

\subsection{local cubic potential}
\label{sec:cubicpot}
  The equation (\ref{eq:grad}) is too general to be very helpful. However, it may describe a saddle-node bifurcation where a stable equilibrium disappears, assuming that $V$ depends slowly on time in a prescribed way, to become a function $V(x,t)$. Near the transition, one may use a
mathematical picture which is correct for a short time around the
transition if the potential $V(x,t)$ is a smooth function (see below for what happens afterwards).

 Assume first that $V(x)$ does not depend explicitly on time and takes the form
\begin{equation}
 V(x) =- (\frac{1}{3} x^3 + b x)
 \mathrm{,}
 \label{eq:sd}
 \end{equation}

 with $b$ real constant (for the moment).

\begin{figure}[htbp]
 \centerline{$\;\;$
  (a)\includegraphics[height=1.5in]{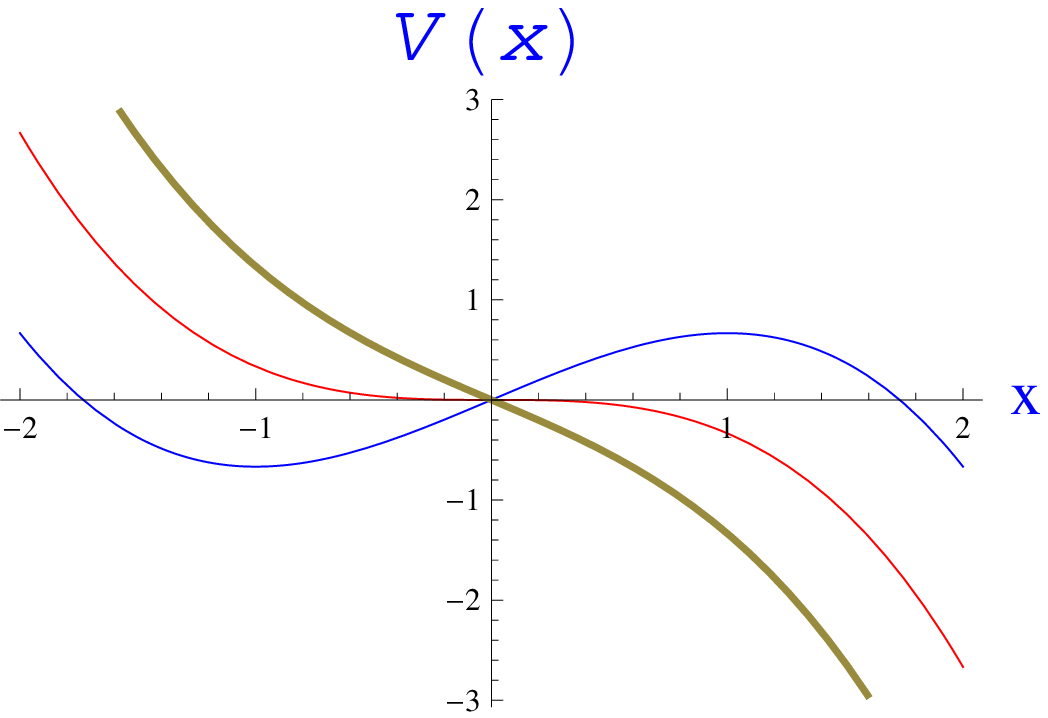}
  (b)\includegraphics[height=1.5in]{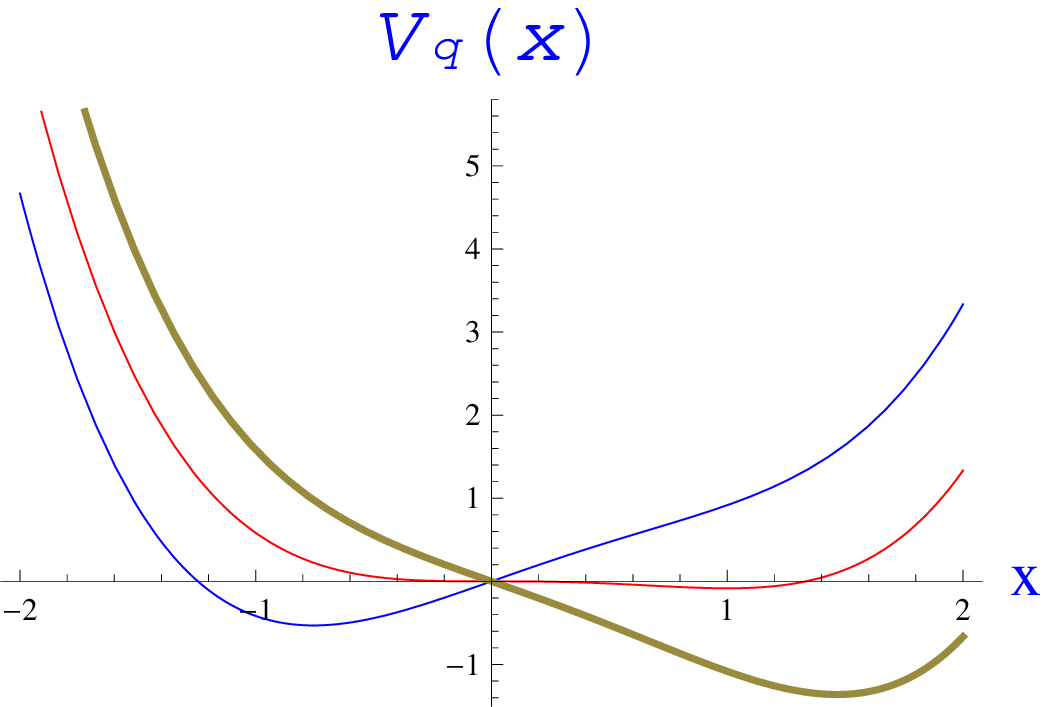}
   $\;\;$}
\caption{(a) Cubic potential for $b =-1,0,1$. (b) Quartic potential,  $ b=-1,0,1$}
\label{Fig:pot}
\end{figure}

For $b$ negative $V(x)$ has two real extrema ({\it{i.e.}} the roots of $\frac{\partial V}{\partial x} =0$), one $-\sqrt{-b}$ is a stable equilibrium, the other, $\sqrt{-b}$, is an unstable equilibrium.
For $b =0$ the two equilibria merge and disappear for $b$ positive, see Fig.\ref{Fig:pot}(a). This is the saddle-node bifurcation. The shape of $V(x)$ near $x =0$ and for $b$ small is universal: for a given smooth $V(x)$ showing this saddle-node bifurcation, one can always rescale $x$ and the external parameter to obtain the "local' problem in this form.

The extension to a time dependent control parameter $b$ goes as follows. If $b$ is a smooth function of time, one can assume that $b(t)$ crosses the critical value, {\it{i.e.}} zero in the present case, at time zero in such a way that $ b(t) = a t +...$ with $a$ non zero constant, dots being for higher terms in the Taylor expansion of $b(t)$. For $t$ and $x$ close to zero, after rescaling, one can represent
 the dynamical system (\ref{eq:grad}), close to the saddle-node bifurcation, by the "universal"
parameterless equation

\begin{equation}
\frac{{\mathrm{d}}x}{{\mathrm{d}}t} = x^2 + t
\mathrm{.}
\label{eq:gradt}
\end{equation}

Outside of the neighborhood of $x=0$, the solution of (\ref{eq:grad}) depends on other parameters defining $V(.)$ for finite values of $x$, as explained below.

Although this is not obvious from its formulation, this model is also valid for the transition from slow to fast motion in the van der Pol oscillator in the limit of large non linearities, as shown by Dorodnitsyn \cite{dor}. Let us sketch the proof of this (interesting) point. Actually we shall look at the formally more general situation of relaxation oscillations, namely at solutions of a set of coupled ODE's (Ordinary differential equations) with a large parameter in the form:
\begin{equation}
 \dot{x} = \eta F(x,y)
 \mathrm{,}
\label{eq:relax1}
\end{equation}
and
 \begin{equation}
 \dot{y} = G(x,y)
 \mathrm{.}
\label{eq:relax2}
\end{equation}
In this set of equations dots are for time derivatives and the functions $F()$ and $G()$ are smooth with values of order $1$ when their argument is also of order one. Moreover, $\eta$ is a large parameter.  The slow manifold is defined by  the condition that, in the limit $\eta$ large, the function $F(x,y)$ must be close to zero over at least part of the trajectory. The Cartesian equation $F(x,y) = 0$ defines a curve in the plane $(x,y)$ which allows to find $y$ as a function of $x$, at least locally. This defines the equation of motion (along the slow manifold) $$ \dot{y} = G(x (y),y) \mathrm{,}$$ where $x(y)$ is such that $F(x(y), y) =0 \mathrm{.}$ The slow trajectory so defined stops at "folds" where the function $x(y)$ ceases to be well defined, namely for values of $(x,y)$ such that $\frac{\partial F}{\partial y} = F_{,y} = 0$, $F(x,y) = 0$,  $F_{,yy}$ being not zero. This defines a discrete set of points. Near those points, the equation of motion can be solved by taking   $\delta x = x - x_0$ and  $\delta y = y - y_0$ where $(x_0 , y_0)$ are the Cartesian coordinates of the point such that

\begin{equation}
F = F_{,x} = 0
 \mathrm{.}
\label{eq:relax4}
\end{equation}
 Let us look at the solution of the coupled equations (\ref{eq:relax1}), (\ref{eq:relax2}) near $(x_0 , y_0)$. The equation  (\ref{eq:relax2})  is not singular at this point and so can be solved for small variation of $y$ and for $t$ small, like $\delta y = t G_0$, $t = 0$ being the (arbitrary) time where the trajectory is in $O$ and $G(x_0,y_0) = G_0$. Consider now the first equation (\ref{eq:relax1}), and expand its right-hand side for $\delta x$ and $\delta y$ small. Because $F_{,x} = 0$ at $(x_0 , y_0)$, the first nontrivial term in the Taylor expansion of $F(.)$ near $(x_0, y_0)$ with respect to $x$ is $\frac{1}{2} (F_{,xx})_0 \delta x^2$. On the other hand the first term coming from the expansion with respect to $\delta y$ is $F_{,y} \delta y = (F_{,y})_0  G_0t$. Therefore, for $t$ and $\delta x$ small, the equation for $\dot \delta x $ derived from  (\ref{eq:relax1}) reads:
\begin{equation}
 \delta\dot{x} = \eta \left( \frac{1}{2} (F_{,xx})_0 \delta x^2 + (F_{,y})_0 G_0 t \right)
 \mathrm{,}
\label{eq:relax3}
\end{equation}
One can check that all terms not written explicitly there are effectively negligible compared to the ones kept. Standard rescalings allows to transform the equation (\ref{eq:relax3}) into the "universal" equation (\ref{eq:gradt}), provided various constraints of sign are satisfied by the quantities $F_{,xx}$,  $F_{,y}$  and $G(.)$ all computed at $(x_0 , y_0)$. This is a short version of the classical calculation by Dorodnycsin, showing that the "universal" equation (\ref{eq:gradt}) is also relevant for the slow to fast transitions in relaxation oscillations. It is worth mentioning also that this derivation does not really assume that the overall motion is periodic, as it may be extended quite easily to a system of many more coupled ODE's with only one fast variable. This shows that such large jumps are also possible in non periodic dynamics, because the dynamics on the slow manifold may be chaotic in between the jumps if this slow manifold has a number of dimensions sufficiently large.

Now we shall focus on the slowing down near the dynamical saddle-node bifurcation described by equation (\ref{eq:gradt}). We shall first give its explicit solution. We look for a solution of equation (\ref{eq:gradt}) transiting from the "stable" fixed point at "large" negative times to the rolling down towards positive value of $x$ at positive times. This  solution behaves like $x(t) \approx - \sqrt{-t}$ at large negative times. The equation (\ref{eq:gradt}) is of the Riccati type and can be integrated by introducing the function $y(t)$ such that $x(t) = -\frac{\dot{y}}{y}$ where $\dot{y} = \frac{{\mathrm{d}}y}{{\mathrm{d}}t}$ and $y(t)$  is a solution of Airy's equation
$\ddot{y} + t y = 0
\mathrm{.}$

The solution of equation (\ref{eq:gradt}) relevant with the given condition at $t\rightarrow-\infty$ is drawn on Figure(\ref{Fig:quartic}-a). In terms of the variable $y(t)$ it is the Airy function  $Ai(-t)$  which writes

 $$Y(t) = Ai(-t)=\int_0^{+\infty} \cos(\frac{u^3}{3} - ut) {\mathrm{d}}u
 \mathrm{.}$$

Yet we have only solved the transient problem near the saddle-node bifurcation.  The transition ends-up when $t$ becomes equal to the first zero of the Airy function $Ai(-t)$, {\it{i.e.}} the smallest root of the equation

$Y(t) = 0$, a pure number, about  $t_c \approx 2.338$. It  corresponds to a divergence of $x(t) = -\frac{y'}{y}$, which behaves as

\begin{equation}
x(t) \approx \frac{1}{t_c - t} -\frac{t_c}{3}(t_c-t)+...
 \mathrm{.}
\label{eq:singdn}
\end{equation}

just before this transition, as derived by expanding $Y(t)$ close to $t_c$.

 Therefore the "generic" equation (\ref{eq:gradt}) for the dynamical saddle-node bifurcation displays a finite time singularity. Let us precise the following mathematical subtlety. This property of the \textit{local flow},
   which results from the folding of the slow manifold, differs qualitatively from the finite time singularity found in the Dieterich-Ruina equations \cite{Chua}, where it was a property of the flow \textit{reduced to the slow manifold} which is everywhere convex, as discussed in the introduction. In the case  of the dynamical saddle-node bifurcation the singularity requires one to consider both the dynamics on and off the slow manifold, and this happens because the geometry forbids the continuation of an exact trajectory on this folded slow manifold.
Actually the solution (\ref{eq:singdn}) loses its physical meaning sometime before the singularity since the "universal" dynamical equation  (\ref{eq:gradt}) was derived under the assumption that $x$ remains close to zero. This local theory cannot deal with finite variations
away from the critical values, therefore we shall need to add finite amplitude effects to limit the growth of the instability after the
transition, see the subsection \ref{sec:quarticpot}.

We shall study now two questions,
first the response of this dynamical system to an external noise,
then the dynamics of a system showing a saddle-node bifurcation of the
type just studied and reaching a new stable fixed point after this
bifurcation. We explore first the response of our system to a small external noise, and look for qualitative changes in
this response
which could be a signal that occurs before the transition.

Let us consider the equation (\ref{eq:gradt}) with a small noise
added, so that equation (\ref{eq:gradt}) is replaced
by

\begin{equation}
\dot{x} = x^2 + t + \epsilon \xi(t)
\mathrm{,}
\label{eq:gradtn}
\end{equation}
where $\xi(t)$ is a random function of time, and $\epsilon$ a small coefficient.

In the limit  $\epsilon$ small, one can solve equation (\ref{eq:gradtn}) by expansion in powers of $\epsilon$,
$ x(t) = x_0 (t) + \epsilon x_1(t) + ...
\mathrm{.}$

 where

 $$ x_0(t) =  -\frac{Y'(t)}{Y(t)}
\mathrm{,}$$

The linear response to the noise is
\begin{equation}
 x_1 (t) = \frac{1}{Y^2(t)} \int_{t_0}^t {\mathrm{d}}\tilde{t} \,\xi(\tilde{t}) \ Y^2(\tilde{t})
  \mathrm{.}
 \label{eq:x1.1}
 \end{equation}
Because $Y^2(\tilde{t})$ tends rapidly to zero as $(\tilde{t})$ tends to minus infinity, one can take $t_0 = -\infty$
to get rid of the effect of the initial conditions.

To make the developments above more concrete, let us take a delta-correlated (or white) noise, such that
$< \xi(t_a) \xi(t_b)> = \delta(t_a - t_b)
\mathrm{.}$

The pair correlation of $x_1(t)$ is given by
$$ <x_1 (t) x_1(t')>  =  \frac{1}{Y^2(t) Y^2(t')}  \int_{-\infty}^{\inf(t,t')} {\mathrm{d}}\tilde{t} Y^4(\tilde{t}) \mathrm{,}$$
where ${\inf(t,t')}$ is the smallest of the two real numbers $t$ and $t'$.
The behavior of this pair correlation
for large negative values of both $t$ and $t'$,
is derived from the asymptotic expression of Airy's function, $Ai(-t) \approx \frac{e^{ - \frac{2}{3} (-t)^{3/2}}}{2\sqrt{\pi}(-t)^{1/4}}$.
Setting $w=\frac{\tilde{t}}{t}$, and $F(w)=1-w^{3/2}$, the variance of the fluctuations writes

$<x_1 (t)^2>  \approx  (-t)  \int_{1}^{\infty} \frac{{\mathrm{d}}w }{w}e^{\frac{8}{3} (-t)^{3/2} F(w)}
 \mathrm{.}$

In the limit $(-t) \rightarrow \infty$ the integral is concentrated near $w = 1$
so that

\begin{equation}
 <x_1 (t)^2>  \approx  \frac{1}{4}(-t)^{-\frac{1}{2}}
 \mathrm{,}
 \label{eq:limcor}
 \end{equation}


which shows that the amplitude of the fluctuations increases some time before the transition itself.
As the transition approaches,
 the standard deviation of the fluctuations,
$ \sigma(t)=\sqrt{<(x(t)-x_0(t))^2>}
 \mathrm{,}$
  increases close to the critical time $t_c$, because $Y(t_c)=0$.

\subsection{Quartic potential}
\label{sec:quarticpot}
   As the zeroth order solution diverges at $t=t_c$, it does not make sense to describe the dynamical behavior of the fluctuations due to the external noise very close to $t_c$, as shown above.

As said before, this unbounded growth of the fluctuations is a consequence of the {\emph{local}} cubic form of $V(x)$ when expanded
near $x =0$,
 in obvious contradiction with the fact that $x(t)$ tends to infinity.
To suppress the divergence of $x(t)$ after
the saddle-node bifurcation we add a stabilizing (positive) term to the potential $V(x)$ which becomes quartic,

\begin{equation}
V_q(x) = - \frac{x^3}{3} - b x + \frac{x^4}{4}
\mathrm{,}
\label{eq:quartic}
\end{equation}

as drawn in Fig.\ref{Fig:pot}(b).
Because of the growth  of $V_q(x)$ at infinity, like $x^4$, the solution of the differential
equation
\begin{equation}
\dot{x} = b + x^2 - x^3
\mathrm{.}
\label{eq:dynquartic}
\end{equation}
does not diverge at finite time. Notice that, formally this does not apply to cases where the local equation (\ref{eq:gradt})  describes a slow-to-fast transition in a limit cycle, instead of a gradient flow dynamics. However, unless one insists to look at what happens well after the saddle-node bifurcation, the details of the dynamics after the fast slide are not significant, because one can consider that the point of landing on the slow manifold after the fast drift is like a new equilibrium for a gradient flow, this neglecting the slow motion on this manifold.

The equation (\ref{eq:dynquartic}) can be written in the given scaled form for any quartic potential
provided the coefficient of $x^4$  is positive. For such a potential
one parameter only remains. In equation
(\ref{eq:quartic}), we choosed to keep as explicit coefficient the coefficient $b$ of the linear term in equation (\ref{eq:quartic}). For $b = 0$ the dynamical system
(\ref{eq:dynquartic})
is exactly at the saddle-node bifurcation, because at $b = x =0$ both the first and second derivative of $V_q(x)$ vanish, but not the third derivative.
Contrary to the case of the pure
cubic potential, this system has always, that is for any value of $b$, a
stable fixed point beyond the pair of fixed points collapsing at the
saddle-node bifurcation. This makes it a fair candidate for describing the
dynamical saddle-node bifurcation without blow-up.

\begin{figure}[htbp]
\centerline{$\;\;$
(a)\includegraphics[height=1.25in]{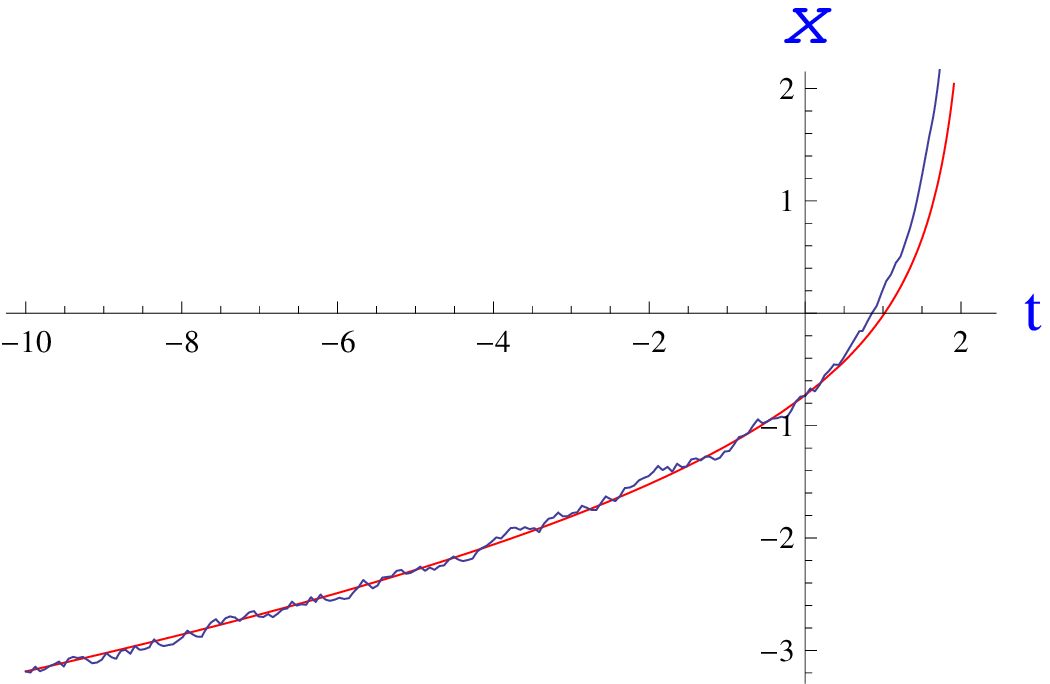}
(b)\includegraphics[height=1.5in]{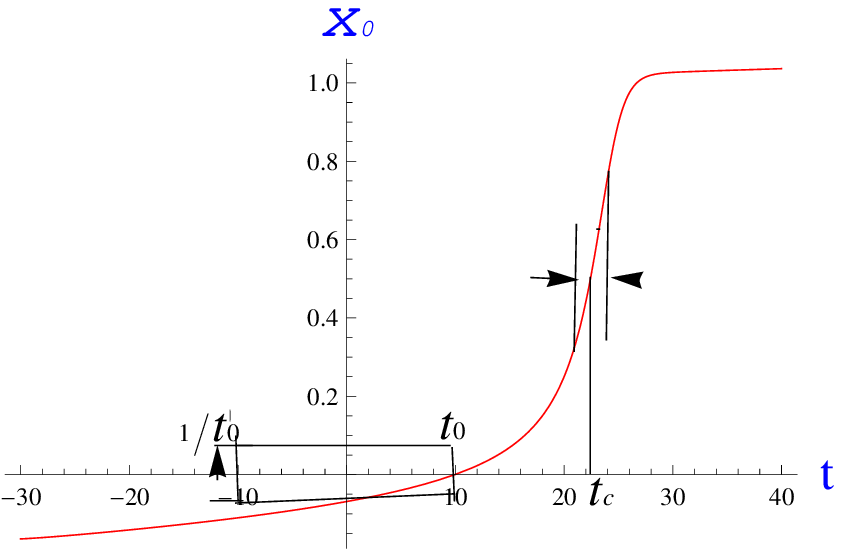}
(c)\includegraphics[height=1.25in]{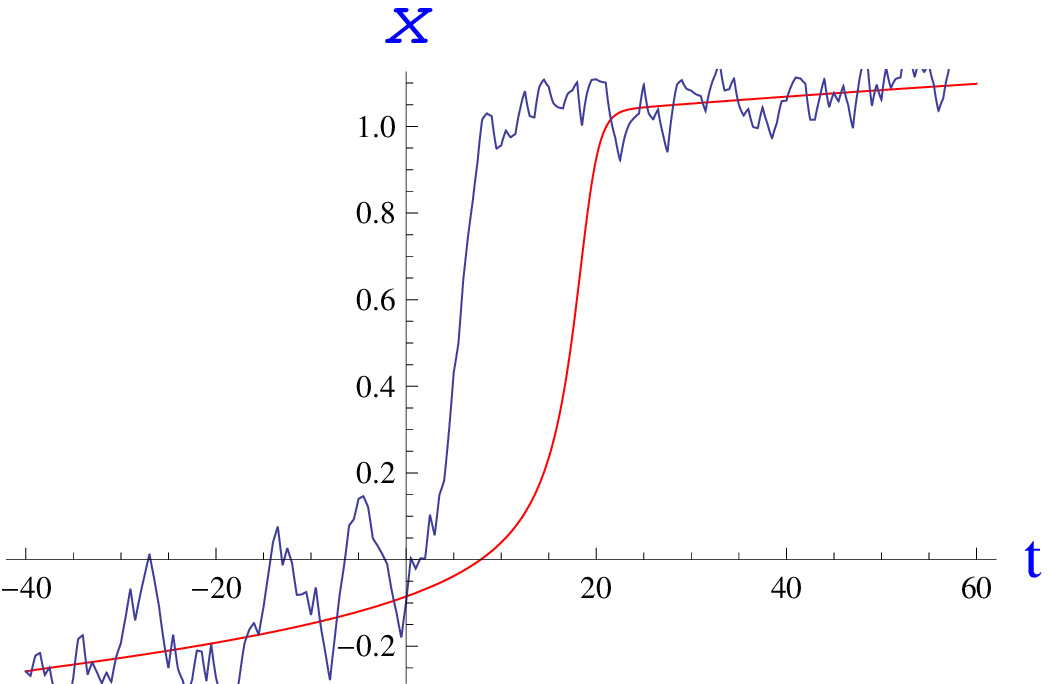}$\;\;$}
\caption{(a) solutions of equation (\ref{eq:gradtn}), with and without noise,
$\epsilon=0$ (smooth curve) and  large amplitude noise $\epsilon=1$.
(b) Solution of equation (\ref{eq:dynquarticscaling})  for $a=10^{-3}$. The rectangle around the origin defines the region  $-t_0 < t <t_0$  and $-1/t_0 < x <1/t_0$, with $t_0\sim a^{-1/3}$. The critical time is $t_c\sim 2.34$ $ t_0$. The two vertical lines inserted between the two arrows delimitate the large slope time duration, of order unity.
(c) Solution of equation (\ref{eq:xxquartic}) with large amplitude noise $\epsilon=1$}.
\label{Fig:quartic}
\end{figure}

 As in the previous case, we shall take now a time dependent $b$, that will be
 taken as $ b = a t$ with $a$ positive constant.
 Because of the rescaling of the cubic and quartic term, the parameter $a$ cannot be eliminated
  (another possibility would be to put a parameter in front of the cubic term).
For the potential $V_q(x) = -\frac{x^3}{3} - a t x + \frac{x^4}{4}$ we shall analyze the solution of the dynamical equation
\begin{equation}
\dot{x} = a t + x^2 - x^3
\mathrm{,}
\label{eq:dynquarticscaling}
\end{equation}
tending at large negative and positive times to the equilibrium point $ x = (a t)^{1/3}$, $t$ being
considered as a parameter, see Fig.\ref{Fig:quartic}(a). Moreover we consider the limit $a$  small,
 $a$ being related to the ratio of small to large time scales, it can be estimated from experimental data. We shall prove that
 in this limit $a$ small, there are three characteristic time intervals, depending how $x$ is
close to zero.

The long time scale is the time lapse between successive major slips, typically of order $150-300 s$ in our experiments on creeping. In our model it is the time needed for the potential $V_q(x,t)$ to
change significantly, to move from a pair of fixed points to a saddle-node
bifurcation. Because time enters in $V_q(x,t)$ through the combination
$(at)$, the a-dimensional time needed for a change of shape of $V_q$ is of order
\begin{equation}
t_{creep} \sim a^{-1}
\mathrm{.}
\label{eq:temps}
\end{equation}

The short time $t_{slip}$ is of order unity in units of our model equation (\ref{eq:dynquarticscaling}) as shown in the next paragraph. It is the duration of the abrupt change of slope of the function of time $x(t)$, the rising of $x(t)$ at $t_c$. In the creep experiment reported above, it is about $1 s$. Therefore the ratio of these two time scales $t_{slip}/t_{creep}$ is as small as $10^{-2}-10^{-3}$ in our experiments.

There is another time scale, $t_0$, the time interval standing before the transition, and close to it, during which the potential is very flat, while the solution  has not yet jumped. During this time, $x$ and $at$  are much smaller than unity, then the cubic term on the right-hand side of equation
(\ref{eq:dynquarticscaling}) is negligible. In this range one recovers the universal
equation of the dynamical saddle-node bifurcation (\ref{eq:gradt}) by
taking $X = x a^{-1/3}$ and $T = t a^{1/3}$, with the boundary condition
$X(t) \approx - \sqrt{-T}$ at $T$ tending to minus infinity. This property concerns the rectangular domain drawn on Fig. \ref{Fig:quartic}-(a), where $x$ is small, $x \sim a^{1/3}$, and  $t$ extends from $-a^{-1/3}$ to $t \sim
a^{-1/3}$, located before the abrupt increase. Therefore the time extension of this domain introduces the intermediate time scale,
\begin{equation}
t_0  \sim a^{-1/3}
\mathrm{,}
\label{eq:temps}
\end{equation}
long compared  to  the short time $t_{slip}$ (which is of order unity, see below)  and small compared to $t_{creep}=a^{-1}$ the average time between slips.

We now show that the short time is of order unity, by matching the solution $X(T)$ of
the universal equation
to the solution of equation (\ref {eq:dynquarticlate}) below, in the vicinity of the critical point $t_c(a)=a^{-1/3} t_c$. Because $X(T)$ behaves like$\frac{1}{t_c-T}$ before it diverges, it follows that the solution
  $x(t)$ behaves as
$\approx \frac{1}{a^{-1/3}t_c- t}$ for "large" values of $\delta
t = a^{-1/3}t_c-t$  before the critical time. This becomes of order one when $\delta
t$ becomes also of order one. When this happens, the
term $at$ in equation
(\ref{eq:dynquarticscaling}) is negligible, therefore the solution of this
equation, which can be matched with the solution near the bifurcation, is
the solution of the integrable equation
\begin{equation}
\dot{x} =  x^2 - x^3
\mathrm{,}
\label{eq:dynquarticlate}
\end{equation}
 with the asymptotic behavior for very large negative times $x(\delta t)
\sim - \frac{1}{\delta t}$.
This equation shows that, in our model,  the fast time scale  is of order one, because it has no explicit dependence with respect to the small parameter $a$. This result is confirmed by the numerics, see Fig.\ref{Fig:quartic}(a). More precisely defining the rising time of $x(t)$ by $1/5$ of the width of the time derivative $\dot{x}$, we get
\begin{equation}
t_{slip}  \sim 1
\mathrm{,}
\label{eq:fastt}
\end{equation}
independently of the value of $a$, for  $a$ small. In the experiment, the intermediate time scale is $t_0^{phys} \sim a^{-1/3}t_{slip}^{phys}$, which gives about $5s$ for $a=10^{-3}$, $t_{slip}^{phys}=1s$.

In summary , by matching the two solutions in the range $1 \ll (-\delta t) \ll a^{-1/3}$, we show that the catastrophe takes place during this time $t_{slip}$ which is of order one, because the displacement is then of order one, compared to the small displacement of order $a^{1/3}$ taking place during time $t_0=a^{-1/3}$ typical of the  "universal" transition process.

From this understanding of the various scales in the deterministic part of the dynamical equations, we can now look at the response to noise of this system, particularly at the range of time where something like a "critical slowing down" could be observed, and which actually happens in our experiments.

 With a noise source added, the dynamical equation (\ref{eq:grad})  becomes,
\begin{equation}
\dot{x} = x^2 - x^3  + a t +\epsilon \xi(t)
\mathrm{.}
\label{eq:xxquartic}
\end{equation}

Actually the effective noise amplitude is not equal to $\epsilon$ close to the saddle-node, as it depends on the value of the parameter $a$. Indeed for $|t|\leq t_{0}$ , the cubic term in equation (\ref{eq:xxquartic}) is negligible, and the equation reduces to
\begin{equation}
\dot{x} = x^2 + a t +\epsilon \xi(t)
\mathrm{.}
\label{eq:xxquartic}
\end{equation}
which may be written as  $\frac{{\mathrm{d}}X}{{\mathrm{d}}T} = X^2 +  T +\tilde{\epsilon}(a) \xi(t)$,
by setting $X=x a^{-1/3}$,  $T=ta^{1/3}$, and $\tilde{\epsilon}(a)=\epsilon a^{-2/3}$.
Therefore the effective noise is larger than $\epsilon$, by a factor $a^{-2/3}$ in the rectangular domain of
 fig.\ref{Fig:quartic}.

\begin{figure}[htbp]
\centerline{$\;\;$
(a)\includegraphics[height=1.6in]{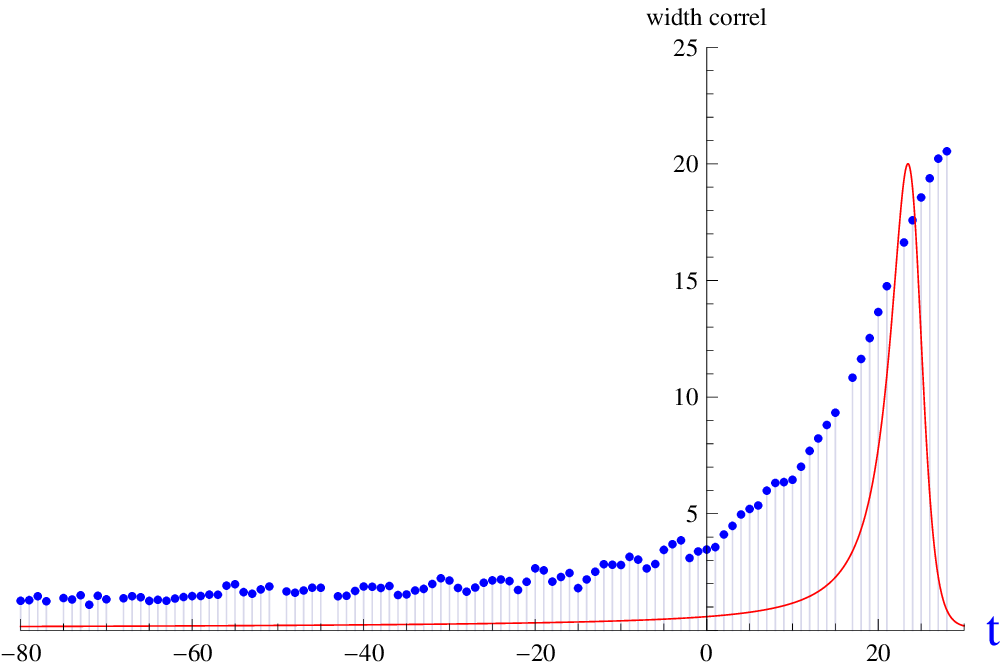}
(b)\includegraphics[height=1.6in]{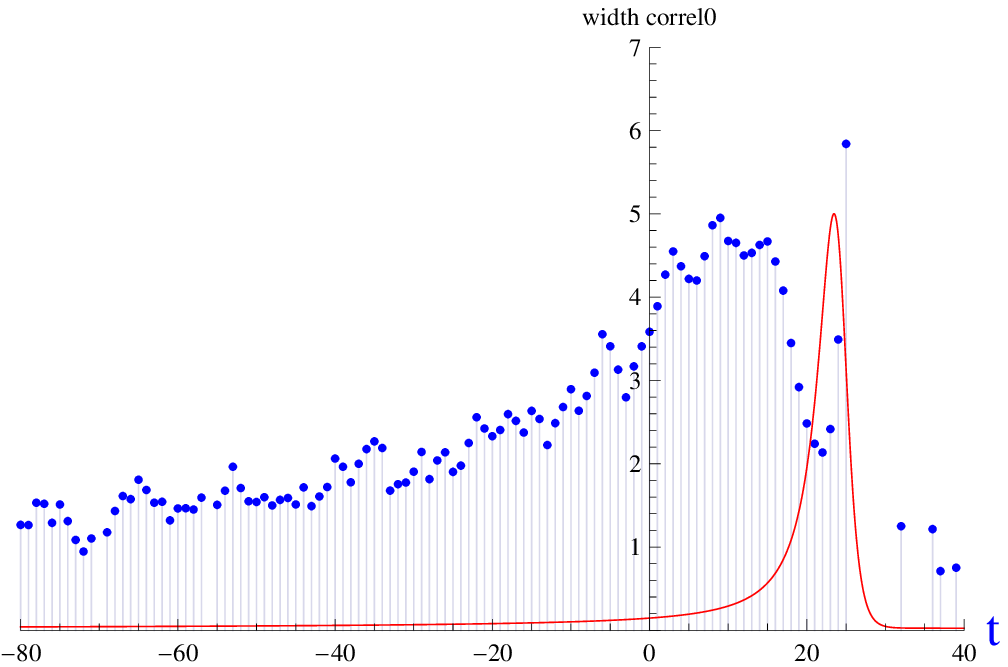}$\;\;$}
\centerline{$\;\;$
(c)\includegraphics[height=1.75in]{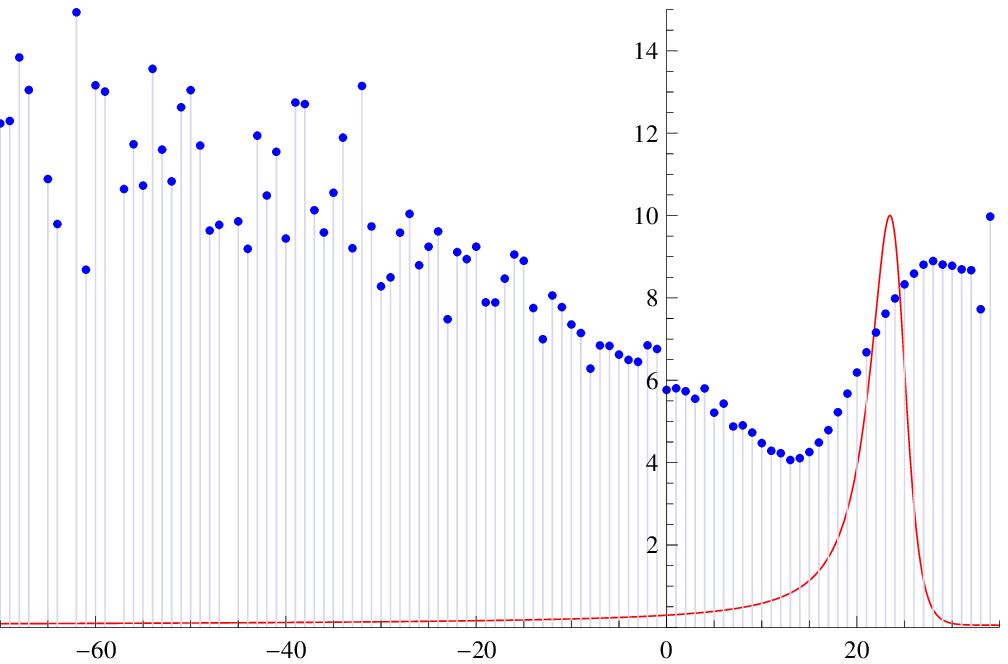}
(d)\includegraphics[height=1.5in]{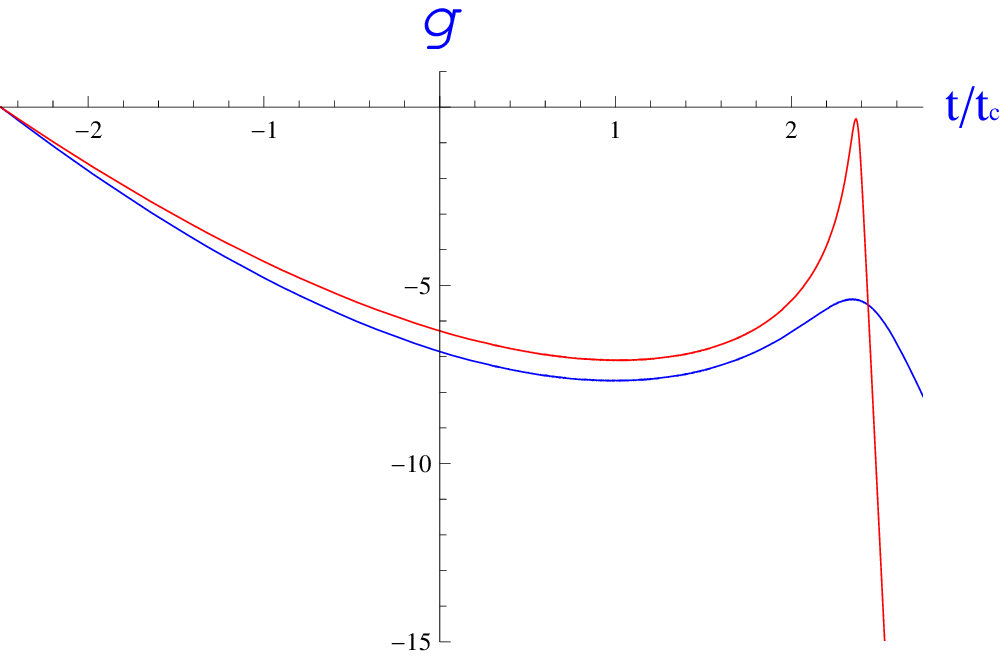}
$\;\;$}
\caption{(a-b) width (in arbitrary units) of the correlation functions of the fluctuation $x(t)-x_0(t)$ for the saddle-node model (\ref{eq:xxquartic}) with $a=10^{-3}$. The correlations functions are defined by equation (\ref{eq:gamma}) and (\ref{eq:gammap}) for curves (a) and (b) respectively; (c) Spectral width (a. u.) calculated with $w_g=t_0$, for $a=10^{-3}$.  (d)  $g(t/t_c)$  for $a=10^{-3}$ (blue),for  $a=10^{-6}$ (red).
}
\label{Fig:width}
\end{figure}

Let us consider now the fluctuations of the solution $x(t)$ of equation (\ref{eq:xxquartic}).
For a small noise source, the solution may be expanded in power of $\epsilon$ as above. At first order it gives
\begin{equation}
\dot{x}_1 = [2x_0(t) -3 x_0^2(t)] x_1(t) + \xi(t)
\mathrm{,}
\label{eq:x1}
\end{equation}
whose solution is formally
\begin{equation}
x_1 (t) = \int_{t_0}^t {\mathrm{d}}\tilde{t} \ \xi(\tilde{t}) \exp[g(t)-g(\tilde{t})]
\mathrm{.}
\label{eq:x1i}
\end{equation}
In general $g(t)$ is the time integral of the second derivative of the potential $-\frac{d^2V_q(x)}{dx^2}$, which yields with our choice of $V(.)$: $$g(t)=\int_{t_0}^t [2x_0(u) -3 x_0^2(u)] \mathrm{.}$$
The standard deviation $\sigma_{x_1}(t)$ has to be calculated numerically. We expect it to display the same behavior as for the cubic case in the whole domain where $x(t) <<1$ , i.e. a little before the transition and close to it, because the potential is cubic in this range. After the transition, we expect that the fluctuation decreases, because the solution without noise becomes quasi-steady. This is confirmed by the numerics: the amplitude of the fluctuations strongly increases close to the critical time $t_c$, its maximum occurring at time $t_c$, then it decreases. More precisely the standard deviation behaves exactly as $\dot{x}_0(t)$, the red curve  in Fig.(\ref{Fig:width}(a), for small noise. Therefore the strong increase of  the variance of the signal fluctuations cannot be used as a precursor for predicting the transition because it occurs simultaneously with the signal itself close to the transition. Note that this observation seems to contradict the currently found statement that fluctuation enhancement precedes the transition and can be used as a precursor. In the case of the saddle-node bifurcation model, we have indeed observed a "precursor" growth of the fluctuations, but only in the case of "large" amplitude noise, see Fig.\ref{Fig:quartic}(b). Let us focus on the case of small noise.
In this case we show below that the correlation time of the fluctuations changes much earlier than the onset of amplitude growth. This slowing-down can be used to foretell the event itself, in considerable advance of its occurrence.  It is seen to consistently happen in the creeping experiment described below.

Consider the case of small effective noise, where the correlation function and the spectrum of the fluctuations $(x(t)-x_0(t))$ are well described by the correlation function and spectrum of $x_1(t)$, respectively. As noted in the previous section, the calculation of these functions requires some care; because the system is not in a statistically steady state. Therefore the
spectral density of the fluctuations depends on time and the correlation function $\Gamma _{x_1}(t,\tau)= <x_1 (t-\tau) x_1(t)>$  depends on both $t$ and on the time difference $\tau$. More precisely, the correct definition of the correlation function is actually given by
\begin{equation}
\Gamma _{x_1}(t,\tau)= \frac{<x_1 (t-\tau) x_1(t)>-<x_1 (t-\tau)>< x_1(t)>}{\sigma_{x_1(t-\tau)}\sigma_{x_1(t)}}
 \mathrm{,}
 \label{eq:gamma}
 \end{equation}
 The latter definition of correlation is not readily accessible in experimental situations, because it requires knowledge of the time dependent variance, which is difficult to estimate from a single sample. A more accessible tool is often used; it
  is given by the expression
 \begin{equation}
\Gamma ' _{x_1}(t,\tau)'= \frac{<x_1 (t-\tau) x_1(t)>-<x_1 (t-\tau)>< x_1(t)>}{\sigma_{x_1(t)^2}}
 \mathrm{,}
 \label{eq:gammap}
 \end{equation}
  which coincides with the correct expression if the variance is the same at time $t$ and $t-\tau$ only. If the variance changes noticeably during the time interval of duration $\tau$, the latter expression  is biased, since $\Gamma=\Gamma' \frac{\sigma(t)}{\sigma(t-\tau)}$. We expect such a discrepancy to manifest itself close to the catastrophe, since the variance increases  there by a large amount. To illustrate this point  we show in Fig.\ref{Fig:width}(a-b)  the correlation functions of the fluctuation $x(t)-x_0(t)$ for the solution of the saddle-node model, as defined by equations (\ref{eq:gamma}) and (\ref{eq:gammap}) respectively. The strong increase of the correlation time before the catastrophe is noticeably truncated when using the biased expression (\ref{eq:gammap}). Using the correct definition of the correlation function (\ref{eq:gamma}), the correlation time increases by a factor of ten over a time interval of order $t_0$ before the catastrophe, while the enhancement is only about $3$ when using the biased expression (\ref{eq:gammap}). Moreover the enhancement is followed by a drop in the latter case. Nevertheless both curves display well the critical slowing down effect, which manifests as a growth of the correlation time close to the transition.  The increase of the correlation length \textit{before} the catastrophe is understood by looking at the formal expression (\ref{eq:x1i}). The second derivative of the potential vanishes at $t=t_0$, and that leads to the flatness of $g(t)$ in the time domain $0 <t <t_c(a)$, as shown in Fig.\ref{Fig:width}(d). This time domain could therefore be identified as a "precursor time", of order few $t_0$.

  Now let us consider the spectrum of the fluctuations in order to compare with the experimental results.
  A time dependent spectrum can be defined formally by the (real) Wigner transform
 \begin{equation}
\mathrm{S} _{x_1}(t,f)= < \int_{-\infty}^{-\infty} {\mathrm{d}}\tau \ <x_1 (t-\tau) x_1(t)> e^{-2i\pi f \tau}
 \mathrm{,}
 \label{eq:spec2}
 \end{equation}
that has to be modified for numerical applications, either by using a filtering procedure like the one used in the previous section or by introducing a slipping window.
In this section we use the latter process. Choosing a Gaussian window function of width $w_g$, the numerical spectrum is given by
 \begin{equation}
\mathrm{S} _{x_1}(t,\nu)= \langle | \int_{t_a}^{t_b}{\mathrm{d}}\tau e^{-(\frac{t- \tau}{w_g})^2}  x_1 (\tau) e^{-2i\pi \nu \tau}|^2 \rangle
 \mathrm{,}
 \label{eq:specn}
 \end{equation}
where $t_{a,b}$ are the numerical integration time boundaries. This expression does not take into account the variability of the variance; therefore we expect the observed change in the spectral properties to be biased, as was the case for the correlation function $\Gamma'$.

 The  evolution of the  spectral width $\Delta f$  (half-height width) is reported in Fig.\ref{Fig:width}(c) in a range of a few times $t_0$ around the transition. The solution ${\dot{x}}_0(t)$ is hightlighted by the solid red line.
  The (c) part of the figure illustrates the same effect as the correlation function $\Gamma'$, but in the Fourier space. The spectral width
decreases noticeably  from  negative time of order few $t_0$, until the time $t\sim 1.5*t_0$, where it grows, because of the bias due to the variance increase close to the burst. The decrease of $\Delta f$ corresponds to a shift of the spectrum  towards low frequencies. The important result is that this shift occurs well \textit{before the transition}.  It occurs over a time interval of order few $t_0$. This result agrees with our experiment where $t_0$ was estimated to be about $10 s$ and the decrease of the spectral width keeps on for about $30s$ ( see Fig.\ref{Fig:creep-exp2} where $10s$ corresponds to $1300$ counts in the abscissa scale).
  The growth of the fluctuations and their shift to lower frequencies can be understood as follows. As the transition approaches, the potential $V(x,t)$ becomes flatter and flatter, making weaker and weaker the restoring force toward equilibrium. Therefore, at constant noise source, the amplitude of the fluctuations driven by this noise source will grow because the damping is ever less efficient. Moreover, the typical time scale for this damping will get increasingly larger because of the decreasing stiffness of the potential, thus favoring noise at lower and lower frequencies.

 \section{Ananthakrishna model}
\label{sec:AK}

This section is devoted to an analysis of solutions of a set of equations devised for describing creeping in solids. More precisely our purpose is to show that, in a range of parameters this equation exhibits a dynamical saddle node bifurcation. As in the ``abstract" model of the previous section, this bifurcation is also preceded by a slowing-down of the fluctuations triggered by an external source of noise.
Creeping phenomena in real materials are complex and difficult to predict quantitatively, despite decades of efforts on theoretical models. We have chosen to consider a model developed by Ananthakrishna for creeping in strained solids. To make things simpler, we have only used its version without space dependence in the quantities involved.
Note that the introduction of space variables would lead to an aperiodic creep signal more realistic than the periodic signals of the present model, however we conjecture that it should not affect the main result of our study (the emphasis of a precursor signal over a given time interval).
 The AK model considered here is a set of three coupled non linear ordinary equations with three dimensionless parameters ($a$, $b$ and $c$, where the letters $a$ and $b$ have no connection with the same symbols used previously). The unknown time dependent quantities
are three scaled
variables corresponding to three density types of dislocations, $x(t)$, $y(t)$ and $z(t)$, representing respectively mobile, immobile, and those with clouds of solute atoms that mimic Cottrell's idea. The model equations write,

\begin{equation}
 b\dot x(t) = G(x,y)= (1-a)x(t)-b x(t)^2-x(t)y(t)+y(t)
  \mathrm{.}
 \label{eq:x1}
 \end{equation}

 \begin{equation}
 \dot y (t) = F(x,y,z)= b x(t)^2-x(t)y(t)-y(t)+az(t)
  \mathrm{.}
 \label{eq:y1}
 \end{equation}

  \begin{equation}
 b \dot z (t) = H(z,x)= c(x(t)-z(t))
  \mathrm{.}
 \label{eq:z1}
 \end{equation}

 where the variable $x(t)$ stands for the elongation
 rate. As one can check, if the three variables are positive at time zero, they remain so at later times if $a$, $b$ and $c$ are positive, as assumed. The relative elongation (or strain, or creep) $L(t)$ of the solder wire is the time integral of $x(t)$, $L(t)= \int _0 ^t x(t')dt'$.
 Solutions of equations (\ref{eq:x1})-(\ref{eq:z1}) have been extensively studied \cite{AnantaR}, their shape and duration versus the parameter values are given in \cite{AK2}. Recall that relaxation oscillations are depicted for small values of the parameter $b$ only. For $c$ larger than a certain critical value $c_{cr}$, depending on $a,b$ and not written here, the solution is stable. The fixed point coordinates are given by the expressions

    \begin{equation}
 \left \{ \begin{array}{l}
x_s=\frac{1-2a}{2b} \, \,  \, \, \,  \, \, \, \, \, \,  \, \, \, \, \, \, \, \, \, \, \, \, \, \, \, \, \, \, \, \, \, \, \, \mathrm{for} \, \, \, \, a< \frac{1}{2}\\
x_s=(1-a)(1+\sqrt{2})\, \,  \, \, \, \mathrm{for} \, \, \, \, a> \frac{1}{2}\\
 y_s=\frac{1}{2}
\mathrm{,}
\end{array}
\right. \label{eq:ptfix}
\end{equation}
 at lowest order for the small parameter $b$. This fixed point becomes unstable by a Poincar\'e-Andronov bifurcation for small values of $b$, only under the condition that $c$ becomes smaller than $c_{cr}$ . In a large range of parameter values the limit cycle associated with the variable $x(t)$ displays relaxation oscillations characterized by slow steps and  fast bursts.

\begin{figure}[htbp]
\centerline{$\;\;$
(a)\includegraphics[height=1.50in]{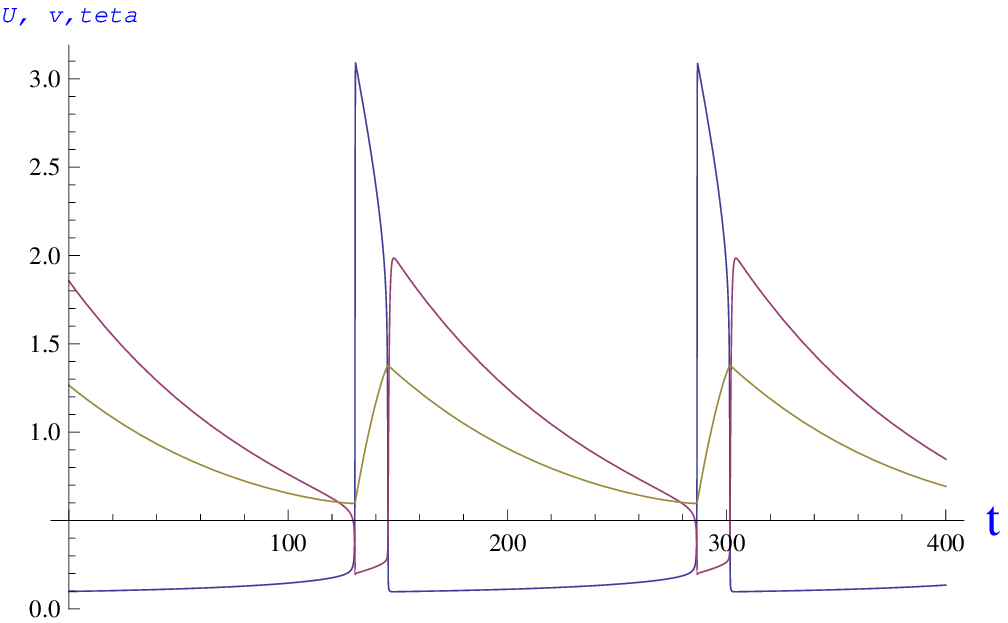}
(b)\includegraphics[height=1.5in]{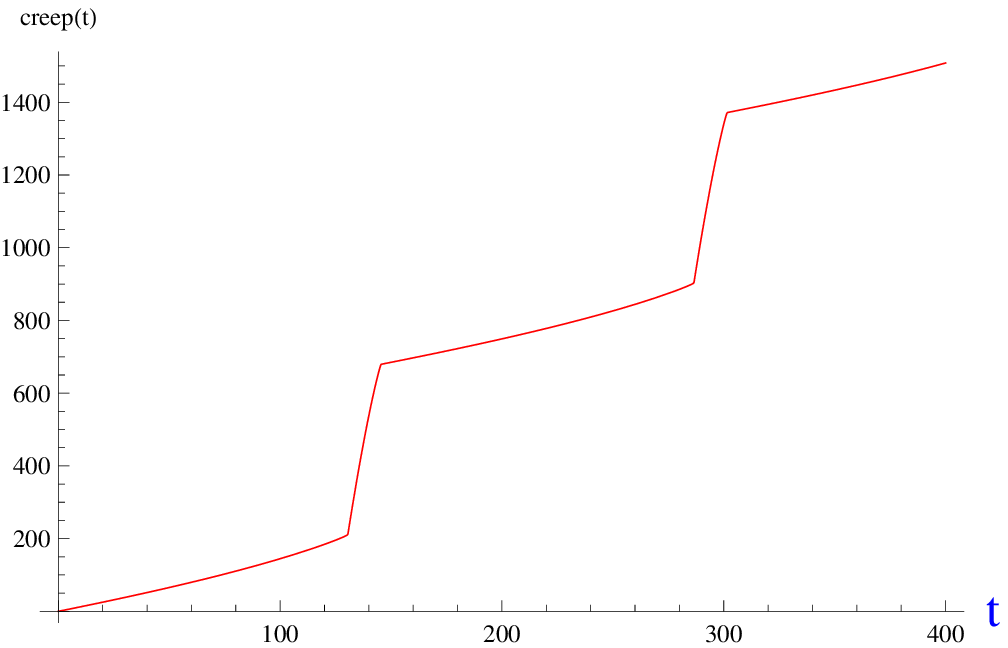}$\;\;$}
\caption{ (a) Solution of the 3D flow(\ref{eq:x1})-(\ref{eq:z1})  for $a=0.65$ , $b=4 10^{-3}$, and  $c=b/100$ . The three curves are  $x(t)/20$ (blue),  $y(t)$ (purple) and $z(t)/8$ (yellow), (b) Creep signal $L(t)$.
 }
\label{Fig:2droite}
\end{figure}

 In this range of
  $b$ and $c$ small, we focus on the case $c<<b$, for two reasons. First because it allows the set of equations to be reduced to the generic saddle-node equation introduced in the previous section. Secondly because it leads to $x(t)$ solutions looking approximately like our experimental data (the strain rate $x(t)$ has to be compared with the data files labeled $y(n)$ in section \ref{sec:experiment}).
   An example of the numerical solution of equations (\ref{eq:x1})-(\ref{eq:z1}) is given in Fig.\ref{Fig:2droite}. In this case the burst amplitude drawn in figure (b)) is noticeably larger than the experimental one shown in Fig.\ref{Fig:creep-exp}-(a), nevertheless we have chosen the parameter values for pedagogical reasons, in order to give a clear representation of the full cycle of the relaxation oscillations. Consider the evolution of the elongation rate $x(t)$. The limit cycle of duration $T=180$ displays four stages. In the first step,   $(0<t<130$, the flow evolves slowly and $x(t)$ takes values of order unity. This stage is followed by a fast jump of $x(t)$ at $t\sim 130$, then by a short  slow stage  $(130<t<144)$ with high values of $x(t)$, of order $1/b$, finally followed by a fast decrease of $x(t)$ at $t\sim 144$.

 \subsection{Stability of the 3D flow}

We first consider the  linear stability  of the 3D flow, especially along the first stage described  just above, which precedes the fast jump. In the next subsection we derive the local form of the model close to the burst. We shall prove that both approaches provide the same information, namely an estimation of the precursor time value. In the slow regime, assuming a 3D flow  of the form  $ x(t)=x_0(t) + \delta x \exp{ \lambda t}$ (and similar expressions for $y(t), z(t)$) close to the trajectory $x_0(t),y_0(t),z_0(t)$, the exponents $\lambda$ are the eigenvalues of the Jacobian matrix
 \begin{equation}
\left(
  \begin{array}{ccc}
   ( 1-a-2bx_0-y_0)/b & (1-x_0)/b & 0 \\
    2bx_0-y_0 & -(x_0+1) & a \\
    c/b & 0 & -c/b \\
  \end{array}
\right)
 \mathrm{,}
  \label{eq:jacobian}
 \end{equation}

or solutions of the equation

\begin{equation}
 \lambda^3+ a_2\lambda^2+a_1\lambda+a_0=0
\mathrm{,}
 \label{eq:lambda3}
 \end{equation}

with $
a_2=(c-(1-a-y_0))/b+3x_0+1$, $a_1=(2x_0+(y_0+a-1)/b)(x_0+1+c/b)+(x_0+1)c/b-(x_0-1)(y_0-2bx_0)/b$, and $a_0=-c(1+x_0)(2bx_0+y_0+a-1)-(y_0-2bx_0)(x_0-1)+a(x-0-1)/b^2$.

\begin{figure}[htbp]
\centerline{$\;\;$
(a)\includegraphics[height=1.50in]{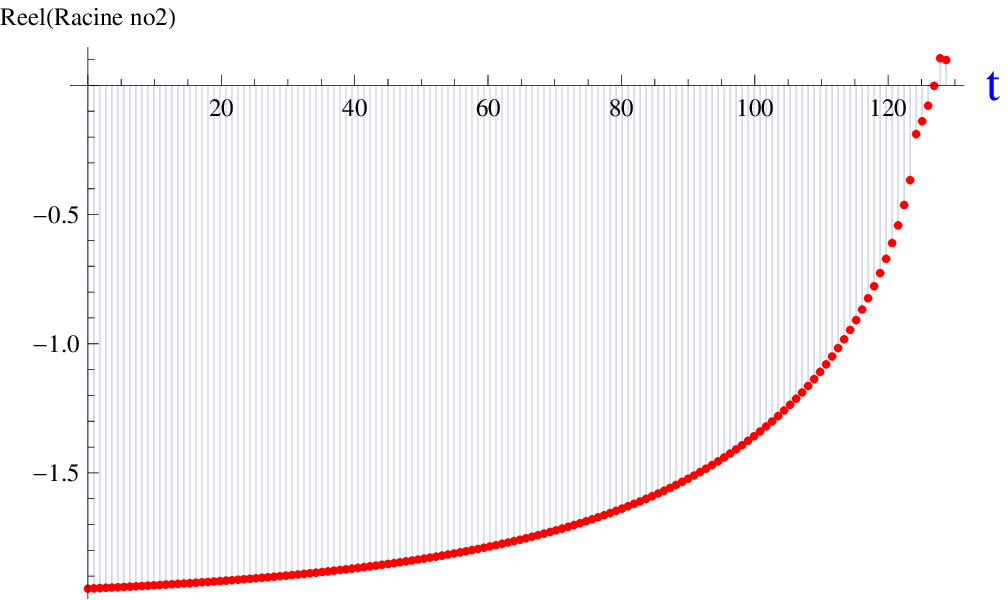}
(b)\includegraphics[height=1.5in]{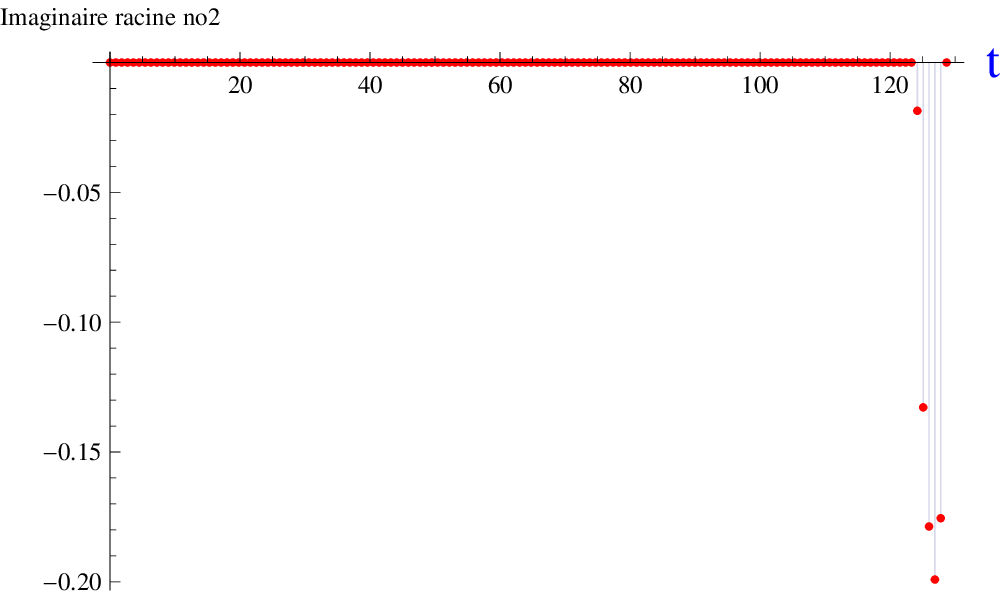}$\;\;$}
\caption{ (a) Real and (b) imaginary part of one of the two eigenvalues crossing the real axis before the burst for the case of the 3D flow illustrated in figure (1).
 }
\label{Fig:lyap}
\end{figure}

One of the eigenvalues is real and negative all along the trajectory, while the other two become complex conjugates in the slow regime, their real part crossing zero before the burst at time $t_{lyap} \sim t_c-3$, see Fig.\ref{Fig:lyap}. The time interval $t_c-t_{lyap}$  depends on the values of the parameters $a,b,c$. It is generally a small fraction of the limit cycle period. Below it is shown that the numerical value of $t_c-t_{lyap}$  is nearly equal to the intermediate time scale $t_0$ for the model.

\subsection{Normal form close to the burst}

We now consider the behavior of the 3D flow in the vicinity of the burst, and derive the normal form of the AK model close to $B$.
 \subsubsection{slow manifold}
 We focus on the first part of the limit cycle, preceding the burst of $x(t)$, and try to understand how the trajectory leaves the slow manifold (SM) that we define now.
Because $c<<b<<1$, the slow stages are described by canceling the  right-hand side of equations
 (\ref{eq:x1})-(\ref{eq:y1}), that reduces the 3D flow to the 1D flow

    \begin{equation}
 \left \{ \begin{array}{l}
G(x,y)=0\\
F(x,y,z)=0\\
 \dot z (t) =  \frac{c}{b}(x(t)-z(t))
\mathrm{.}
\end{array}
\right.
\label{eq:sm}
\end{equation}

\begin{figure}[htbp]
\centerline{
\includegraphics[height=1.50in]{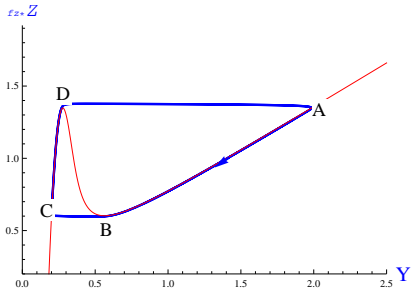}}
\caption{ SM $z(y)$ (red curve) for $a=0.65$ , $b=4 .10^{-3}$, and parametric plot of the 3D flow (blue curve, numerical solution $z(x(y(t)),y(t))$ of equations (\ref{eq:x1})-(\ref{eq:z1}) for $c=b/100$.
}
\label{Fig:nullcline}
\end{figure}

From the first equation the variables $x(t)$ is an explicit function of $y(t)$,
  \begin{equation}
x(y) = \frac{-(y-1+a)+\sqrt{(y-1+a)^2+4by}}{2b}
  \mathrm{.}
 \label{eq:xs}
 \end{equation}
Inserting this expression into the second equation (\ref{eq:sm}), allows to also express the variable $z(t)$ in terms of $y(t)$,

   \begin{equation}
z(y) = \frac{-bx(y)^2 +y(x(y)+1)}{a}
  \mathrm{,}
 \label{eq:zs}
 \end{equation}
In the phase space $(z,y)$ expression (\ref{eq:zs}) defines the slow manifold, which is illustrated in Fig. (\ref{Fig:nullcline}), red curve.
On the positive slope parts of the slow manifold  the 1D flow obeys the differential equation $$ \dot y (t) = \frac{c}{b} \frac{1}{z_{,y}} (x(y(t))-z(y(t)))$$, which becomes singular at critical points defined (on the SM) by the relation
\begin{equation}
 z_{,y}=0
  \mathrm{.}
 \label{eq:dF}
 \end{equation}

 The 3D flow (closed blue curve) follows the path $A \rightarrow B \rightarrow C \rightarrow D$. At the critical point $B$, the trajectory leaves the slow manifold, jumps to point $C$  (fast stage $B \rightarrow C$), then it follows the portion ($C \rightarrow D$), and finally returns to the SM in $A$.
We consider below the exit of the SM close to  the critical point $B$.

\subsubsection{Critical point B}

The coordinates of the critical point $B$ in the phase space $x,y,z$ are solutions of equations (\ref{eq:xs}) and (\ref{eq:dF}).
 From equation (\ref{eq:xs}) we derive an expression for $ x_{,y}=\frac{\partial x}{\partial y}$
  \begin{equation}
 x_{,y}= -\frac{(x-1)^2}{1-a+bx^2-2bx}
  \mathrm{.}
 \label{eq:F1}
 \end{equation}

 which should be identical to the expression of $x_{,y}$ taken from the relation (\ref{eq:dF}),
  \begin{equation}
(x_{,y})_B= \frac{x^2-1}{3bx^2-(2b+1-a)x}
  \mathrm{.}
 \label{eq:F2}
 \end{equation}

 These two expressions are identical if $x_B$ is a root of the cubic polynomial equation
  \begin{equation}
 x(x-1)(1+2\tilde{b}-3\tilde{b}x)= (x+1)(1-2\tilde{b}x+\tilde{b}x^2)
  \mathrm{.}
 \label{eq:xB}
 \end{equation}
 where $\tilde{b}=b/(1-a)$. The positive solution $x_B$ and the corresponding value of $y_B=((1-a)x_B-bx_B^2)/(x_B-1)$ and $z_B$ satisfying equations (\ref{eq:xs})-(\ref{eq:zs}) respectively are given at leading order with respect to the small parameter $\tilde{b}$ by

   \begin{equation}
 \left \{ \begin{array}{l}
x_B=\sqrt{2}+1 + \frac{8+5\sqrt{2}}{2}\frac{b}{1-a}\\
y_B=(1-a)(1+\sqrt{2})-\frac{8+11\sqrt{2}}{2}\frac{b}{1-a}\\
z_B=\frac{(3+2\sqrt{2})(1-a)}{a}-b \frac{10+7\sqrt{2}}{a}
\mathrm{,}
\end{array}
\right.
\label{eq:turn}
\end{equation}

valid  in the parameter range $b<<(1-a)$.

\begin{figure}[htbp]
\centerline{$\;\;$
(a)\includegraphics[height=1.250in]{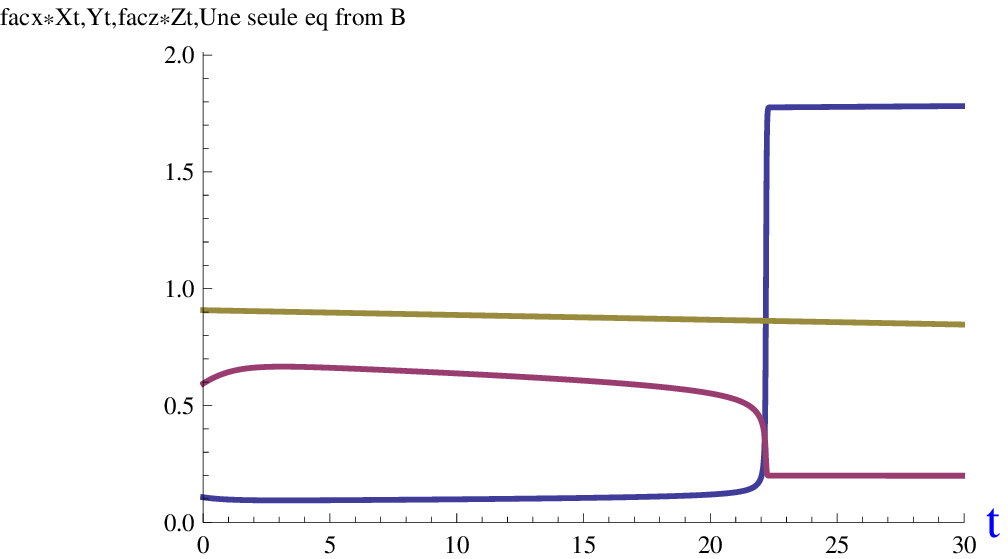}
(b)\includegraphics[height=1.5in]{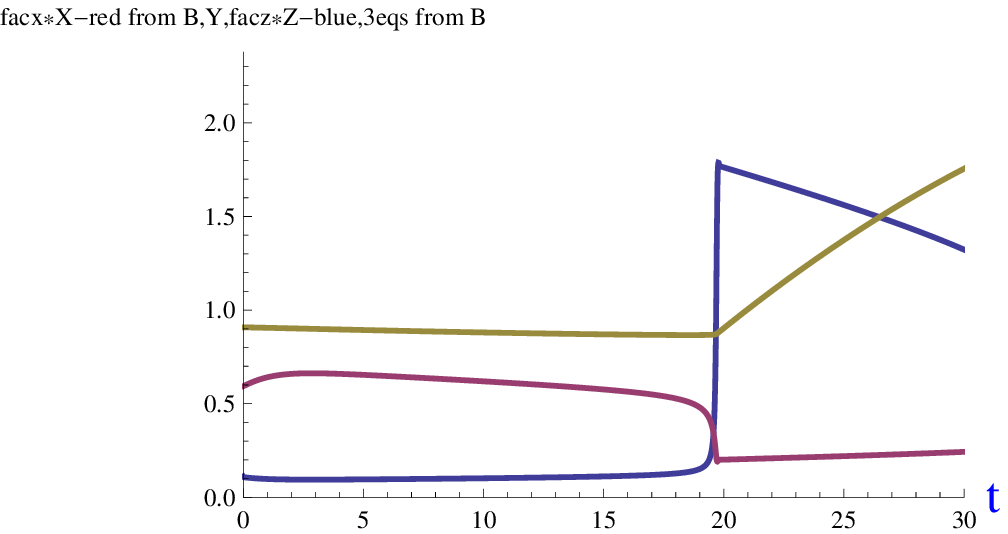}$\;\;$}
\caption{ Solution from point B  to point C, (a) Solution of the 1D flow equation (\ref{eq:z1}) with relations (\ref{eq:xs})-(\ref{eq:zs}), (b) solution of the 3D flow equations (\ref{eq:x1})-(\ref{eq:z1}).
}
\label{Fig:closeT}
\end{figure}

We next derive the normal form describing the dynamical behavior of the solution close to the critical point. Assuming that the variable $x$ follows adiabatically the variable $y$ according to equation (\ref{eq:xs}), the original system reduces to
   \begin{equation}
 \left \{ \begin{array}{l}
\dot{y}=F(y,z)\\
\dot{z}=\frac{c}{b}(x(y)-z)
\mathrm{,}
\end{array}
\right.
\label{eq:2eqB}
\end{equation}
defining the dynamics for the portion  of the trajectory near $B$, including the burst $B \rightarrow C$. This is confirmed by the numerics: starting from the point B , with numerical initial conditions (\ref{eq:turn}), the behavior of the solutions of equation (\ref{eq:2eqB}) agree well  with the 3D flow, see Fig. \ref{Fig:closeT}.

Since we are interested in the description of the solution before and at the burst, in the intermediate regime where $x$ remains close to  $x_B$ , we shall pursue our analysis by canceling the terms $bx^2$ in (\ref{eq:xs}) and in $F$, because $bx_B<<1$. This leads to the system

   \begin{equation}
 \left \{ \begin{array}{l}
 x=\frac{y}{y-1+a}\\
\dot{y}=\tilde{F}(y,z)\\
\dot{z}=\frac{c}{b}(x(y)-z)
\mathrm{,}
\end{array}
\right.
\label{eq:2eq}
\end{equation}
where $\tilde{F}=-xy-y+az$. Close to $B$ and during the burst, the variable $z$ is essentially constant, due to the small value of the ratio $c/b$, whereas the variable $y$ jumps toward smaller values, as shown in Fig. \ref{Fig:nullcline}.
Therefore at leading order with respect to the small parameter $c/b$, the solution of the last equation (\ref{eq:2eqB}) is given by

 \begin{equation}
z ( t) =z_B+ \frac{c}{b}(x_B-z_B) t
 \mathrm{.}
\label{eq:zbt}
\end{equation}
in the vicinity of $B$.
 Inserting this local solution (\ref{eq:zbt}) for $z(t)$ into the system (\ref{eq:2eq}), the exit from the SM is then described by a single equation  for the local variation $\delta y= y-y_B$ of $y$ of the variable $y$ , which is of the form

 \begin{equation}
{\delta  \dot {y}} = \sum_{n>2} \frac{1}{n!}(\tilde{F}_{,y^n})_B \delta y^n  +\gamma \delta t
 \mathrm{,}
\label{eq:s1}
\end{equation}

 where
  \begin{equation}
\gamma=(ac/b)(x_B-z_B)
 \mathrm{,}
\label{eq:gamma}
\end{equation}

 and $\tilde{F}^{(n)}_B$ sets for the $n^{th}$ derivative of $\tilde{F}$ with respect to the variable $y$, taken at point $B$ ( note that 
$(F_{,y})_B=0$  at the critical point $B$). Using the reduced system (\ref{eq:2eq}), for the first two derivatives of $\tilde{F}$ with respect to $y$ we write
  \begin{equation}
 \left \{ \begin{array}{l}
\tilde{F}_{,y^2}=-yx_{,y^2} -2x_{,y} \\
\tilde{F}_{,y^3}=-y x_{,y^3}- 3 x_{,y^2}
\mathrm{.}
\end{array}
\right. \label{eq:derF1}
\end{equation}

Expanding all expressions close to $B$ at leading order with respect to the small parameter $b$, one obtains
  \begin{equation}
 \left \{ \begin{array}{l}
(x_{,y})_B=-\frac{2}{1-a}\\
(x_{,y^2})_B=-\frac{4\sqrt{2}}{(1-a)^2}\\
(\tilde{F}_{,y^2})_B=-\frac{4\sqrt{2}}{1-a}\\
(\tilde{F}_{,y^3})_B=\frac{24}{(1-a)^2}
\mathrm{.}
\end{array}
\right. \label{eq:derF2}
\end{equation}

At this stage we can limit the series expansion in  equation (\ref{eq:s1}).
A rough approximation for the series to converge is given by $ (\tilde{F}_{,y^3})_B \delta y < 3 (\tilde{F}_{,y^2})_B $. Using expressions (\ref{eq:derF2}) and the first equation (\ref{eq:2eq}) gives the range of variation

  \begin{equation}
 \left \{ \begin{array}{l}
  |\delta y |<\frac{3}{\sqrt{2}}(1-a)\\
  |\delta x |<\sqrt{2}
\mathrm{.}
\end{array}
\right. \label{eq:dxdy}
\end{equation}

In this range
 the local form of the AK equations close to $B$ becomes
 \begin{equation}
 \delta \dot{y} =  \frac{1}{2}(\tilde{F}_{,y^2})_B  \delta y^2  +\gamma \delta t
  \mathrm{.}
 \label{eq:generic}
 \end{equation}
which is identical to equation (\ref{eq:relax3}).

Finally , setting $T=(\frac{-\gamma (\tilde{F}_{,y^2})_B }{2})^{1/3}\delta t $, and $Y=[\frac{((\tilde{F}_{,y^2})_B )^2}{4\gamma}]^{1/3} \delta y$, the relation (\ref{eq:generic}) takes the generic form (\ref{eq:gradt}) without any parameter recently proposed as a possible description of a signal before a catastrophe \cite{nous-saddle}.
   \begin{equation}
\dot{Y} =-Y^2- T
  \mathrm{.}
 \label{eq:scgeneric}
 \end{equation}

As shown in section \ref{sec:dynsaddbif} the solution of  equation (\ref{eq:generic}) displays an intermediate time scale of order
    \begin{equation}
t_0 =(\frac{2}{\gamma(\tilde{F}_{,y^2})_B })^{1/3}
  \mathrm{.}
 \label{eq:scgeneric}
 \end{equation}
 which could be used to predict the burst, because the critical slowing down effect occurs during this time interval.
In the AK equations the  intermediate time scale, or precursor time, is given by the relation

 \begin{equation}
t_0 \sim (\frac{1-a}{2\sqrt{2}\gamma })^{1/3}
 \mathrm{,}
\label{eq:prect}
\end{equation}
where the parameter $\gamma$ is given by (\ref{eq:gamma}).
 Recall that $t_0=t_c - t_{Lyap}$ was shown to be  the time delay between the catastrophe and the time $t_{Lyap}$ at which the instability builds up along the slow manifold. For the parameter values of the above  Figures, we have $t_0 \sim 3.3$, which agrees well with the Lyapunov analysis, see Fig. \ref{Fig:lyap}, and also with the precursor time deduced from the spectral analysis presented in the next subsection.

In summary we have proven that the AK model, although formally different from  the van der Pol model,
has a
 normal form close to the critical point that is consistent with the dynamical
 saddle-mode model equation studied in (\cite{nous-saddle}), with an intermediate time scale given by expression (\ref{eq:prect}).
 Consequently, as for the saddle-node model,  the AK model should
 display a response to noise with a strong increase of the correlation
 time occurring with a few times $t_0$ before the burst.

\subsection{Response to noise before B}

The response to noise of the system (\ref{eq:x1})-(\ref{eq:z1}) is studied by setting ($x=x_0(t)+u_b(t)$, $ y(t)=y_0+v_b(t)$, $z(t)=z_0(t)+ \theta_b(t)$), where $x_0(t),y_0(t),z_0(t)$ is the solution of the noiseless  AK equations, and the vector $V(t)= u_b(t),v_b(t),\theta_b(t)$ characterizes the fluctuations of the response to a noise source. These fluctuations result from the introduction  of noise terms (either multiplicative or additive) in the original system. In the case of  multiplicative noise sources, the response of the  AK equations is a solution of the system

\begin{equation}
 \left \{ \begin{array}{l}
  \dot x(t) =(1/b) [(1-a)x(t)-b x(t)^2-x(t)y(t)+y(t)](1+\epsilon_x f_x(t))\\
 \dot y (t) = [b x(t)^2-x(t)y(t)-y(t)+az(t)](1 +\epsilon_y f_y(t))\\
 \dot z (t) = (c/b)[x(t)-z(t)](1 +\epsilon_z f_z(t))
 \mathrm{.}
\end{array}
\right. \label{eq:eqnoise}
\end{equation}

\begin{figure}[htbp]
\centerline{$\;\;$
(a)\includegraphics[height=1.25in]{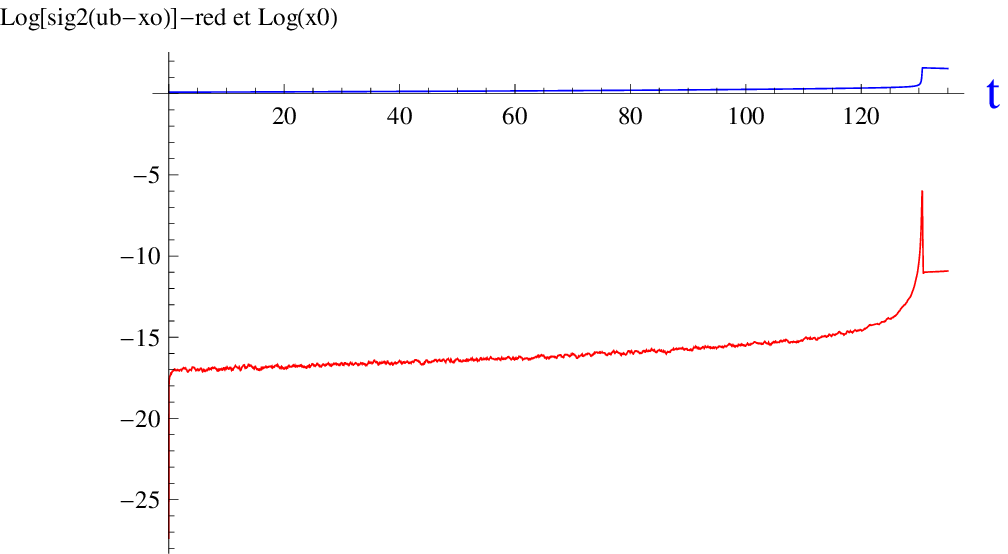}
(b)\includegraphics[height=1.25in]{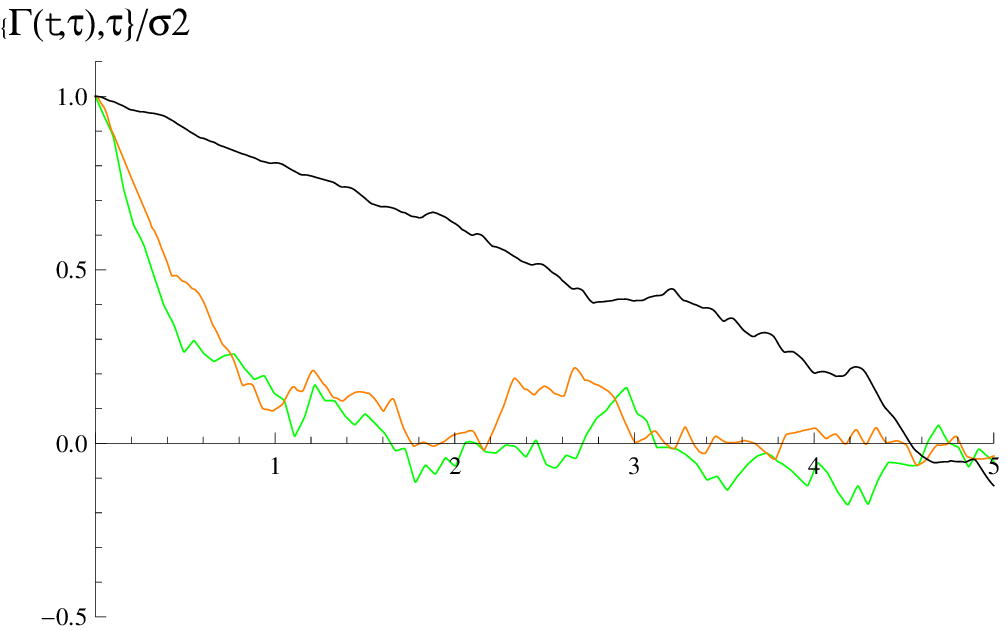}
$\;\;$}
\centerline{$\;\;$
(c)\includegraphics[height=1.50in]{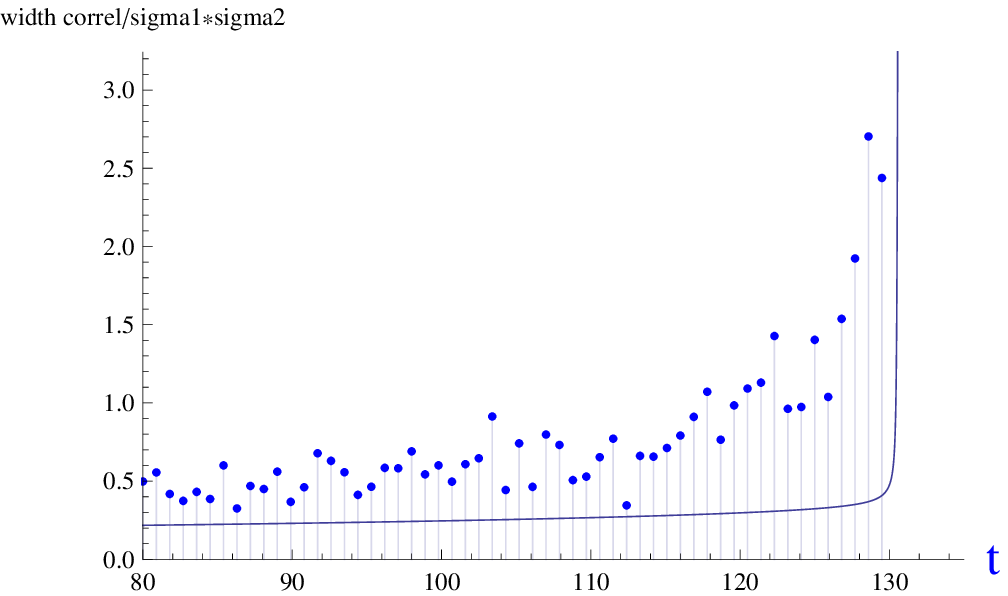}
$\;\;$}
\caption{Multiplicative noise: (a) variance of the response  (red curve) as a function of time, compared to $x_0$ (blue curve),  in Log scale. (b) Correlation functions $\Gamma(t,\tau)$ at times
$t=50$ -green , $t=100$-orange and $t=129$-black curve versus $\tau$. (c) Half-height-half-with of the correlation function $\Gamma (t,\tau)$ versus time $t$. The input data are those of Figure (\ref{Fig:2droite}), noise amplitudes are $\epsilon_i=10^{-5}$ for the three independent noise sources $f_i(t)$.
}
\label{Fig:correl}
\end{figure}

 We report below the result of the numerical study of the correlation function given by expressions (\ref{eq:gamma}) and (\ref{eq:gammap}) for small amplitude noise. In this case the variance increases slowly along the SM, following approximately the evolution of the variable $x_0(t)$, as reported in Fig. \ref{Fig:correl} (a). Figures (b-c) displays  the evolution of the correlation function before the burst, along the path $B\rightarrow C$. The width of the correlation function $\Gamma$ clearly increases as the burst is approached, showing an evident critical slowing down effect. The maximum growth of the correlation time occurs close to the time $t \sim t_{lyap}$ where the SM becomes linearly unstable, persisting until the burst. This result is in agreement with the experimental result given in section \ref{sec:experiment} showing a shift of the spectrum towards low frequencies before the burst.

\begin{figure}[htbp]
\centerline{$\;\;$
(a)\includegraphics[height=1.250in]{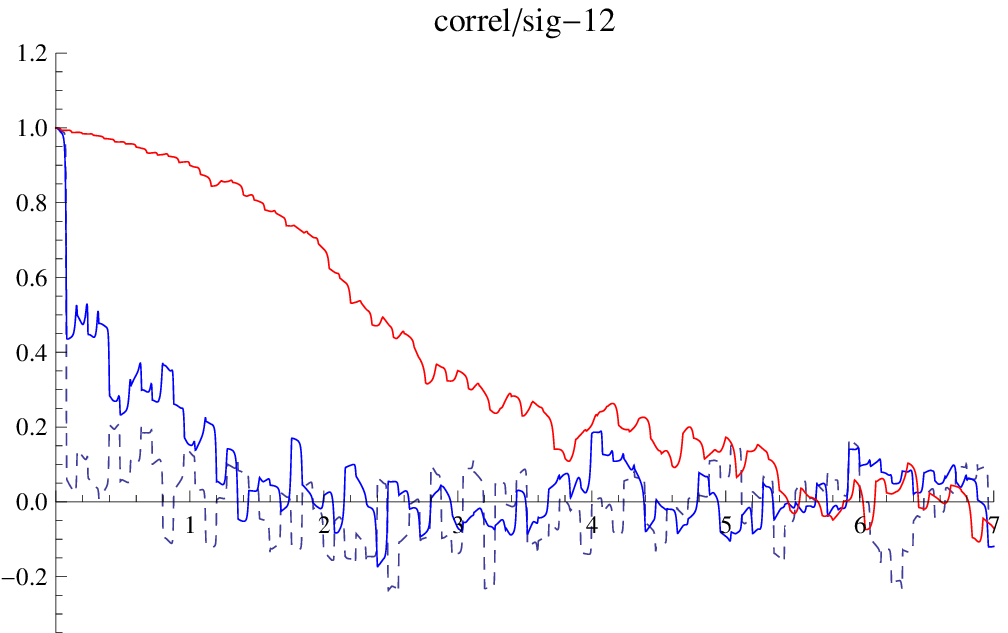}
(b)\includegraphics[height=1.250in]{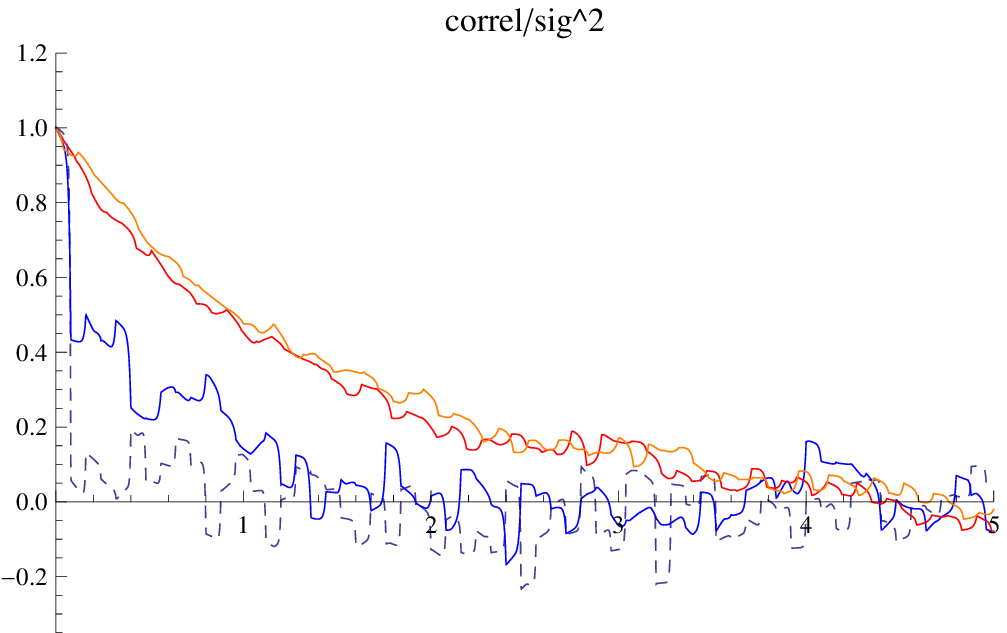}
$\;\;$}
\centerline{$\;\;$
(c)\includegraphics[height=1.25in]{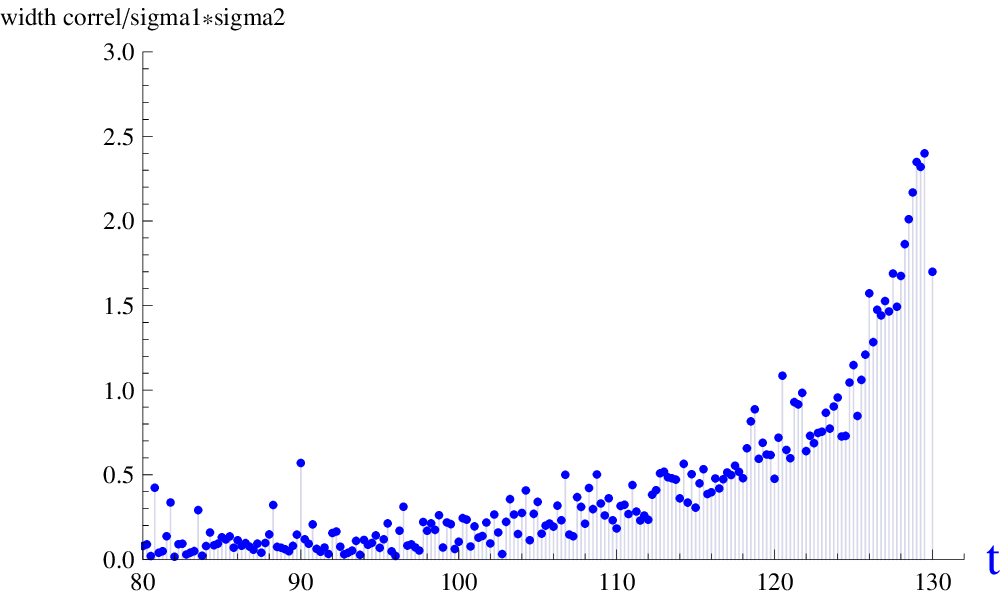}
(d)\includegraphics[height=1.25in]{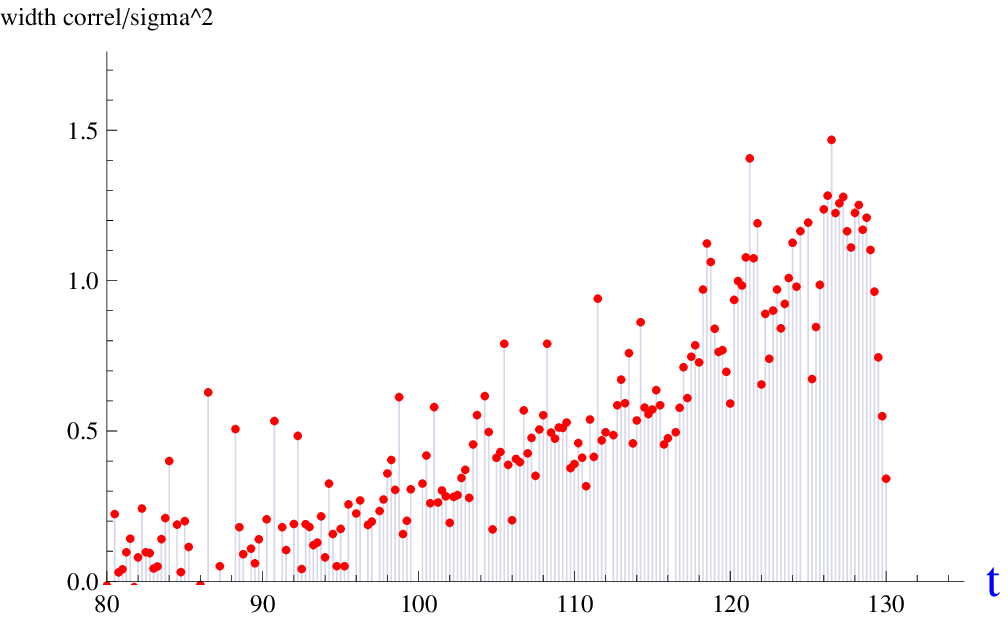}
(e)\includegraphics[height=1.in]{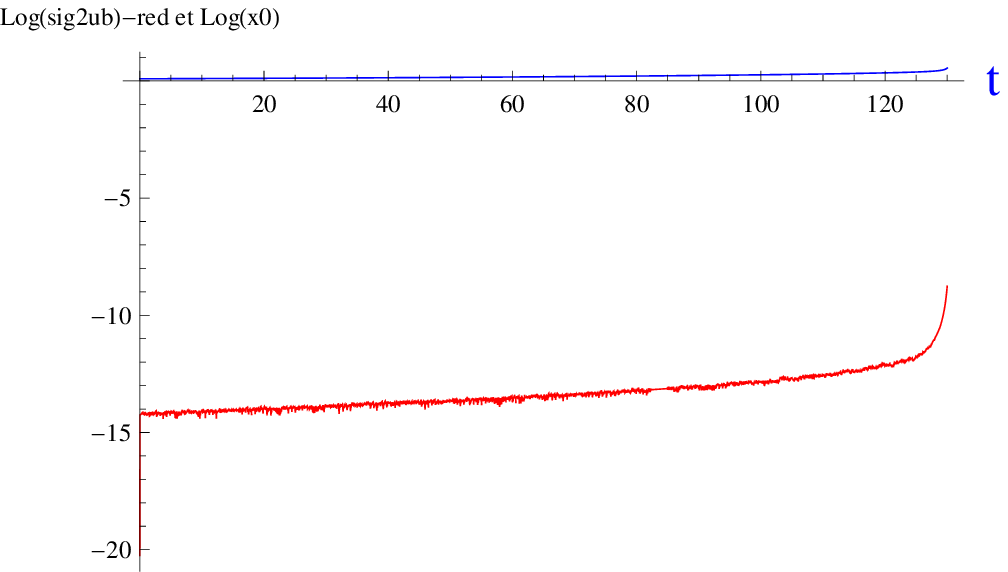}
$\;\;$}
\caption{
 With additive noise. (a-b) Correlation functions at times $t=50,100,129$ (dashed, Blue, Red curves respectively), (a) $\Gamma(t,\tau)$, (b) $\Gamma'(t,\tau)$ versus $\tau$.
(c-d) Evolution of the half-width of the correlation functions  along the trajectory before the burst, (c)  half-width of $\Gamma(t,\tau)$ (d) half-width of $\Gamma'(t,\tau) $ both curves drawn versus $t$.  Same parameters and input data as in Figures (\ref{Fig:correl}). Noise amplitudes are $\epsilon_i=10^{-4},10^{-6},10^{-7}$ for the three independent noise sources $f_i(t)$. (e) Evolution of the variance along the path ($B \rightarrow C$)
}
\label{Fig:correladd}
\end{figure}

 For noise of additive type the width of the correlation function $\Gamma(t,\tau)$ behaves similarly, as shown in Fig. \ref{Fig:correladd}.
As pointed out in a previous section, using the expression (\ref{eq:gammap}) to calculate the correlation time of the solution
  gives a biased result close to the burst where the  variance is time dependent.
In this case the  increase of the width  before the burst is reduced as illustrated in Fig. \ref{Fig:correladd}(b,d).

\begin{figure}[htbp]
\centerline{
\includegraphics[height=1.50in]{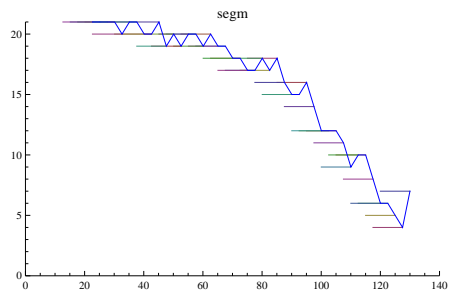}}
\caption{Cumulative spectral width $\omega$ of the response to additive noise $u_b(t)$, versus time before the burst, same  data as in other figures .
}
\label{Fig:specum}
\end{figure}

In order to present the AK model results by using the same analytical tools as used for the  experiment, we have calculated the cumulative spectrum of the response $u_b(t)$,
$\textrm{C}_S(t,\nu)= \int_{0}^{\nu}{d\nu' S(t,\nu')}$
where $S(t,\nu)= <|\int_{t-\Delta t}^{t}{dt' u_b(t')\exp ^{2i\pi \nu t'}}|^2>$ ,
 is the spectrum of the fluctuations for the response signal sampled during the time interval $(t-\Delta t, t)$, in analogy with equations (\ref{eq:spec})-(\ref{eq:cumul}). The characteristic spectral width $\omega$ defined by the relation (\ref{eq:cumwidth}) evolves in time as illustrated
  in Fig. \ref{Fig:specum} for the  AK model with additive noise. The horizontal segments corresponding to abscissa ($t-\Delta t, t$), have ordinates $\omega$, the  numerical value of the spectral width calculated during this time interval. The decrease of $\omega$ close to the burst shows well the expected shift of the spectrum towards low-frequencies, observed also in the experiments.

In conclusion the AK model displays a range of parameter values where one can see the critical slowing-down observed also in the experiments. We have studied the Ananthakrishna model for the case of small $b$ and $c$ parameter values, with $c\ll b$. In this case the burst occurs close to a critical point, as noticed in \cite{AK2}. Close to this critical point, the 3D flow (namely a set of {\emph{three}} coupled ODE's)  can be reduced to the 2D system (\ref{eq:2eq}) with a sort of Langevin-like source term (noise term with small amplitude noise)of the form (\ref{eq:relax1})-(\ref{eq:relax2}) discussed in section \ref{sec:dynsaddbif}. The present system
differs from
the van der Pol equation; however, close to the critical point
both models take the form of the dynamical saddle-mode model equation studied in \cite{nous-saddle}.

The Ananthakrishna model displays a  response to noise with a strong increase of the correlation time occurring close to the burst, or a shift of the spectrum towards low-frequency components. For this model the ``precursor time" is a time interval of order several times  $t_c-t_{Lyap}$, also equal to several times the intermediate time scale  which depends on the  values of the parameters $a,b,c$ as given by equation (\ref{eq:scgeneric}).
A detailed quantitative comparison between the AK model and the creep experiment will require a more complete study, which is beyond the scope of the present work. In particular it should be noted that including spatial variables in the AK model leads to chaotic solutions, which is in better agreement with our experiment where the limit cycle period may indeed vary by a factor of $3$ from one recorded data set to another.

  \section{Conclusion}
 We have shown that the dynamical model of saddle-node transitions recently proposed to foretell catastrophes is applicable to describe the physics of collective dislocations. The  experimental signal of the plastic deformation of the eutectic mixture of Sn-Pb subjected to a constant stress presented above clearly displays the well-known critical slowing down effect, with a precursor time of order $1/10$ of the relaxation oscillation period. This observation is shown to agree with the response to noise of a physical model proposed by   Ananthakrishna for describing creeping in ductile materials. This physical model has  $3$ parameters, therefore it would be quasi-impossible to make a quantitative comparison between the experimental data and the theoretical model. Here we show the AK model has a range of parameter values for which the slow-fast transition is described by the dynamical saddle-node model mentioned above. When adding a small noise to the AK model, we show that it indeed displays the critical slowing down scenario, and we derive the expression for the precursor time in terms of the $3$ parameter values. This gives a relation between the parameter values of the AK model in which the  precursor time (i.e. the time during which the most intense spectral components clearly shift toward low frequencies) is of order one tenth of the relaxation oscillation period, as observed in the experiment. The important result is the following: for such systems which display relaxation oscillations with saddle-node transition,  we show that there exists an intermediate time scale between the slow and fast regimes,  which can be used to foretell catastrophes. Unlike the majority of precursor tools that have focused on signals in the time domain, we have looked at fluctuations in the frequency domain.  These spectral precursors are prone to fewer false alarms.


\begin{thebibliography}{let1}
\bibitem{convex} In situations similar to one of the van der Pol equation in the limit of large nonlinearities, the slow to fast transition occurs at the point on the slow manifold where the large component of the velocity is tangent to the slow manifold. This tangency never happens in the DR equations because, there, the large component of he velocity crosses everywhere the slow manifold at finite angle.  Therefore the scenario of slow to fast transition is different in the DR equations than what it is in limit cycles of the Li\'enard-van der Pol like equations.
\bibitem{apparatus1} This product came from Tel-Atomic, Inc., a company that specializes in ``Tools for Teaching Advanced Physics". It is no longer available as a commercial product. One reason this instrument is no longer manufactured was the loss, due to a factory fire, of readily available
chips that constituted the key component
(NE5521n integrated circuit) used in the support electronics for its displacement sensor\cite{apparatus2}.  A same first example of fully differential
capacitive sensors is also employed (except employed as a pair, for mechanical common mode rejection of noise) with different electronics,
in another of the company's instruments, the
computerized Cavendish balance \cite{apparatus3}.
\end{thebibliography}
\end{document}